\documentclass[12pt]{article}
\usepackage{epsf}
\newcommand{\PSbox}[3]{\mbox{\rule{0in}{#3}\includegraphics{#1}\hspace{#2}}}
\newcommand{\beq}{\begin{eqnarray}}
\newcommand{\eeq}{\end{eqnarray}}
\newcommand{ \centeron }[2]{{\setbox0=\hbox{#1}\setbox1=\hbox{#2}\ifdim
                             \wd1>\wd0\kern.5\wd1\kern-.5\wd0\fi \copy0
                             \kern-.5\wd0\kern-.5\wd1\copy1\ifdim\wd0>\wd1
                             \kern.5\wd0\kern-.5\wd1\fi}}
\newcommand{ \ltap }{\>\centeron{\raise.35ex\hbox{$<$}}
                     {\lower.65ex\hbox{$\sim$}}\>}
\newcommand{ \gtap }{\>\centeron{\raise.35ex\hbox{$>$}}
                     {\lower.65ex\hbox{$\sim$}}\>}

\newcommand{ \lsim }{\mathrel{\ltap}}

\newcommand{ \vv }{u}

\newcommand{\drawsquare}[2]{\hbox{%
\rule{#2pt}{#1pt}\hskip-#2pt
\rule{#1pt}{#2pt}\hskip-#1pt
\rule[#1pt]{#1pt}{#2pt}}\rule[#1pt]{#2pt}{#2pt}\hskip-#2pt
\rule{#2pt}{#1pt}}

\def\ltap{\raisebox{-.4ex}{\rlap{$\sim$}} \raisebox{.4ex}{$<$}}
\def\gtap{\raisebox{-.4ex}{\rlap{$\sim$}} \raisebox{.4ex}{$>$}}
\newcommand{\Yfund}{\raisebox{-.5pt}{\drawsquare{6.5}{0.4}}}

\def\lesssim{\mathrel{\mathpalette\vereq<}}

\makeatletter
\def\vereq#1#2{\lower3pt\vbox{\baselineskip1.5pt \lineskip1.5pt
\ialign{$\m@th#1\hfill##\hfil$\crcr#2\crcr\sim\crcr}}}
\makeatother

\begin{document}

\begin{titlepage}

\begin{center}
\vspace*{-1cm}
~ \hfill   SCIPP-00-27 \\
~\hfill hep-th/0008151 \\
\vskip .3in
{\Large \bf Radion Dynamics and Electroweak Physics}

\vskip 0.15in
{\bf Csaba Cs\'aki$^{a,}$\footnote{J. Robert Oppenheimer fellow.}, 
Michael L. Graesser$^b$ and Graham D. Kribs$^{c,}$\footnote{Address
after 1 September 2000:  Department of Physics, University of Wisconsin,
Madison, WI 53706.}}

\vskip 0.15in

$^a${\em Theory Division T-8, Los Alamos National Laboratory \\
Los Alamos, NM 87545}

\vskip 0.1in

$^b${\em Department of Physics\\
University of California, Santa Cruz, CA 95064}

\vskip 0.1in

$^c${\em Department of Physics\\
Carnegie Mellon University, Pittsburgh, PA 15213}

\vskip 0.1in
{\tt csaki@lanl.gov, graesser@scipp.ucsc.edu, kribs@pheno.physics.wisc.edu}

\end{center}

\vskip .15in
\begin{abstract} 
The dynamics of a stabilized radion in the Randall-Sundrum model (RS) 
with two branes is investigated, and the effects of the radion on
electroweak precision observables are evaluated.  The radius is 
assumed to be stabilized using a bulk scalar field as suggested by 
Goldberger and Wise.  First the mass and the wavefunction of the 
radion is determined including the backreaction of the bulk stabilization 
field on the metric, giving a typical radion mass of order the weak scale.
This is demonstrated by a perturbative computation of the radion 
wavefunction.  A consequence of the background configuration for 
the scalar field is that after including the backreaction  
the Kaluza--Klein (KK) states of the bulk scalars couple directly 
to the Standard Model fields on the TeV brane.  Some cosmological 
implications are discussed, and in particular it is found that the 
shift in the radion 
at late times is in agreement with the four--dimensional 
effective theory result. 
The effect of the 
radion on the oblique parameters is evaluated using an effective 
theory approach.  In the absence of a curvature--scalar Higgs mixing 
operator, these corrections are small and give a negative contribution 
to $S$.  In the presence of such a mixing operator, however, the 
corrections can be sizable due to the modified Higgs and radion couplings.

\end{abstract}
\end{titlepage}

\newpage
\section{Introduction}
\setcounter{equation}{0}
\setcounter{footnote}{0}

Extra dimensional theories where standard model fields are localized on a
brane 
\cite{bow,otherextra,HW,rs1,Gogberashvili,otherRS,inflating,CohenKaplan}
have recently attracted a lot of attention, since such models 
have several distinct features from ordinary Kaluza-Klein (KK) theories.
In particular, Randall and Sundrum \cite{rs1} presented a simple model based 
on two branes and a single extra dimension, where the hierarchy problem
could be solved due to the exponentially changing metric along the extra
dimension. In order to obtain a phenomenologically acceptable model, the
radion field (which corresponds to fluctuations in the distance of the
two branes) has to get a mass, otherwise it would violate the equivalence 
principle \cite{equivalence}, 
and also result in unconventional cosmological expansion
equations \cite{BDL,CGKT}. 
The simplest mechanism for radius stabilization has been
suggested by Goldberger and Wise \cite{gw}, 
who employed an additional bulk scalar
which has a bulk mass term and also couples to both branes
(for another issues related to the radion potential
see \cite{otherstable}).
Both the cosmology and collider phenomenology 
crucially depend on the mass and couplings of the radion. 
In particular, there is no radion moduli problem if the radion 
mass is $O({\rm TeV})$ and its couplings to Standard Model (SM) 
fields is $O({\rm TeV}^{-1})$. This is also the most favorable 
scenario for discovering the radion at a future collider. 

In fact it was shown in \cite{CGRT,GW3}, that the radion will 
have the above properties for the Goldberger--Wise scenario. 
However, the calculation of \cite{CGRT,GW3} were using a na\"{\i}ve 
ansatz for the radion field which ignores both the radion wavefunction
and the backreaction of the stabilizing scalar field on the metric. 
The validity of this approximation has recently been questioned 
\cite{Cline}.
 
Therefore in this paper
we analyze the coupled radion-scalar system in detail from
the 5D point of view. We derive the coupled differential equations 
governing the dynamics of the system, and find the mass eigenvalues
for some limiting cases. Due to the coupling between 
the radion and the bulk scalar, we find that there will 
be a single KK tower describing
the system, with the metric perturbations non-vanishing for every 
KK mode. This implies that the Standard
Model fields localized on the TeV brane will couple to every KK mode from 
the bulk scalar, and this could provide a means to directly 
probe the stabilizing physics. 

Using the coupled equations for the radion--scalar system, we analyze 
the late--time behavior of the radion in an expanding universe, and 
find that the troubling 55--component of Einstein's equation 
just determines the shift of the radion. This shift completely 
agrees with the shift obtained in \cite{CGRT} 
using the 4D effective theory. 
  
Given that we have established that the radion mass is 
$O({\rm TeV})$ and that its couplings to SM particles is 
$O({\rm TeV}^{-1})$, it is reasonable to 
consider its effects on SM phenomenology. Some direct 
collider signatures for the radion and loop corrections have been 
discussed in \cite{rspheno,DHR-big,DHRg-2,Indian,GRW,Korean,Kim}.
In the second half of this paper the effects of the radion 
on the oblique parameters\footnote{Loop effects for theories with
large extra dimensions have been analyzed in \cite{largeextradloops}.} 
are calculated using an effective theory 
approach similar to \cite{effective}. 
Since in the RS model the radion is the only new state 
well below the TeV scale, a low--energy effective theory
including only the radion and SM fields is used. The effects 
of other heavy modes are accounted for by including 
non--renormalizable operators at the cutoff scale.
In the absence of a curvature--scalar Higgs mixing operator, 
the corrections from the radion are small, but give a negative contribution 
to $S$. In the presence of such a mixing operator the corrections
could be sizable due to the modified radion and Higgs couplings.
 
This paper is organized as follows: in Section \ref{review-RS-GW-sec} 
we review the Randall-Sundrum
model and radius stabilization by bulk scalar fields. We also summarize the
explicit example of \cite{dewolfe} which we will be using for our explicit
computation of the radion mass and couplings to SM fields. In Section
\ref{coupled-sec} we present our ansatz for the coupled metric 
and scalar fluctuations
based on the analysis of \cite{CGR,PRZ} of the radion without a 
stabilizing potential. We will derive a single ordinary differential 
equation, whose eigenmodes will yield the KK modes for the radion-scalar
system.  In Section \ref{general-properties-sec} we analyze the 
generic properties of this equation.
In the general case we find that the system is not described 
by a hermitian Schr\"odinger operator.
However, we identify a convenient limit, in which the differential
operator is in fact hermitian, and the eigenfunctions are
manifestly orthogonal. 
In Section \ref{approx-KK-sec} we analyze the eigenfunctions 
in this limit, and find the
approximate masses for the KK tower. In this analysis, the back reaction of 
the metric is neglected, which results in the lightest mode still being 
massless. The effect of the back reaction on the lightest mode is
taken into account in Section \ref{mass-sec}, where we find the mass 
of the radion
to be of the order (but slightly lighter) than the weak scale.
In Section \ref{coupling-KK-sec} we discuss the couplings of 
the radion and the KK tower
to SM fields on the brane. We find that the radion coupling is
{\it exactly} agrees with the results in \cite{CGRT,GW3}, while
the couplings of the other KK modes of the scalar field are suppressed 
by the mass of the given mode, and is proportional to the backreaction
of the metric due to the scalar background.  
In Section \ref{cosmology-sec} we demonstrate that in an 
expanding universe the shift in the radion at late times 
agrees with the 4D effective theory result obtained in 
\cite{CGRT}. 
Having established the mass and coupling of the radion, 
we write an effective Lagrangian in Section \ref{effective-lag-sec}
without any specific mechanism of radius stabilization and
neglecting the contributions of the KK modes.  In Section 
\ref{curvature-scalar-sec} we add a curvature-scalar Higgs 
coupling to the effective Lagrangian, and discuss how the
couplings are modified.  Then, in Section \ref{feynman-sec}
we calculate the Feynman rules in a general gauge.  These
allow us to compute the oblique parameters via one-loop vacuum 
polarization diagrams with radions in the loop in Section
\ref{electroweak-sec}.  The radion correction is log divergent
(unlike the Higgs), and so we also write the nonrenormalizable 
operators at the cutoff scale that provide the necessary counterterms.  
The size of the new contributions are shown for various cases in 
several figures in Section \ref{numerical-sec}.  We also estimate 
limits on the radion mass as a function of the cutoff scale in 
Section \ref{limits-sec}.  Finally, we conclude in 
Section \ref{conclusions-sec}.

\section{Review of the Randall-Sundrum Model and the \\
Gold\-ber\-ger-Wise
Mechanism}
\setcounter{equation}{0}
\setcounter{footnote}{0}
\label{review-RS-GW-sec}

Randall and Sundrum presented a very interesting proposal 
for solving the hierarchy problem \cite{rs1}. 
By introducing a fifth dimension where the bulk geometry 
is anti-de Sitter, a large hierarchy between the Planck 
scale and the TeV scale is obtained with only a mild 
finetuning. Two branes are introduced, located at the 
boundaries of the anti-de Sitter space. By tuning the 
bulk cosmological constant $\Lambda\equiv - 6 k^2 /\kappa^2$,  
the tensions $V_P$ and $V_T$ on the Planck and TeV branes, 
respectively, such that $V_P=-V_T=6 k /\kappa^2$ (where $\kappa^2$ is
the 5D Newton constant related to the 5D Planck mass by $\kappa^2=1/2M^3$) 
one obtains 
a 4-D Poincare invariant solution. The metric is then 
\begin{equation} 
ds^2 =e^{-2ky} \eta _{\mu \nu} d x^{\mu} dx ^{\nu} -dy^2, 
\end{equation} 
where the Planck brane and TeV branes are located at $y=0$ and 
$y=r_0$. 
For a moderate choice of $k r_0 \sim O(50)$,  
a large hierarchy between the Planck 
scale and the weak scale is generated. 

Since this solution is obtained for any value of $r_0$, 
some mechanism is 
required to 
fix $r_0 \sim 50/k$ as opposed to some other value of 
$r_0$. This must also be done without introducing any large 
finetuning. Further, 
small shifts in the separation between the two branes do not 
change the energy, and so 
 are described 
in an effective theory by the fluctuations of a massless particle, 
the ``radion''. This particle couples like a Brans-Dicke scalar and 
must be massive to recover ordinary 4-D Einstein gravity 
\cite{equivalence,CGRT}
 
One way to achieve these requirements is to introduce a bulk 
scalar field $\phi$ that  
has a bulk potential $V(\phi)$ \cite{gw}. To stabilize the 
brane distance, potentials $\lambda_{P,T}(\phi)$ 
on the Planck and TeV branes respectively are also included.
The competition between the brane and bulk Lagrangians
generates a vacuum expectation value (vev) for $\phi$, 
which results in a 4-D vacuum energy that depends on $r_0$. 
For a simple choice of polynomial potentials a large 
hierarchy is then easily obtained with a mild finetuning 
\cite{gw}, and the resulting mass 
for the radion is $O({\rm TeV})$ \cite{CGRT,GW3}. 

The phenomenology of the radion depends on the 
strength of its coupling to the brane fields. Using the following 
na\"{\i}ve {\em ansatz} to describe the radion $b(x)$
\begin{equation} 
ds^2 = e^{-2k |y| b(x)} ds^2_4 - b(x)^2 dy^2 ,
\end{equation} 
\cite{CGRT} and \cite{GW3} computed the normalization of the 
radion kinetic term to be 
\begin{equation} 
\frac{3}{4} e^{-2 k r_0} \frac{k r_0^2}{\kappa^2} 
  (\partial b)^2 .
\label{rkterm}
\end{equation} 
Fields living on the TeV brane couple to the radion through the 
induced metric, with an interaction  
\begin{equation} 
 \frac{k r_0}{2} b(x) \hbox{Tr} T_{\mu \nu}
= \frac{r(x)}{\sqrt{6} \Lambda_W} 
\hbox{Tr} T_{\mu \nu}
\label{rcoupling}
\end{equation} 
where $r$ is the canonically normalized radion, $\Lambda_W 
= M_{Pl} e^{-kr_0} \sim O({\rm TeV})$, $M^2_{Pl}=
(1-e^{-2kr_0})/ (k \kappa^2)\sim 1/ (k \kappa^2)$, 
and $T_{\mu \nu}$ is 
the physical energy-momentum tensor of the TeV brane fields.   
It is then clear that the radion 
couples as $\sim 1/$TeV  
to the Standard Model fields. 
Obtaining an acceptable phenomenology then requires that the 
radion mass is $O($TeV$)$, which is easily satisfied by 
the Goldberger--Wise mechanism.  

The phenomenology of the radion is then crucially dependent 
on the normalization of the kinetic term. In fact, 
in the computation leading to the $O($TeV$)^2$ 
 prefactor in (\ref{rkterm}) there is a cancellation 
between two terms of $O(M_{Pl}^2)$. The origin of this 
cancellation remains somewhat mysterious, and the absence 
of this cancellation would clearly lead to different 
predictions. In \cite{Cline} it was pointed out that there 
are additional contributions to the radion kinetic 
term not included in \cite{CGRT,GW3}. In particular, 
the profile of the stabilizing field depends on $r_0$, and 
so a small change in $r_0 \rightarrow b(x)$ distorts the background field. 
It was found that this results in an 
$O(M_{Pl}^2)$ correction to the radion kinetic term, thereby 
drastically changing the phenomenology of the radion. 

We review the resolution of this issue in the first part of this paper.
Some of the results presented in Sections \ref{coupled-sec}--\ref{mass-sec} 
are already contained in the 
work by Tanaka and Montes \cite{TanakaMontes}, even though the results
of this paper were obtained independently of Ref.\cite{TanakaMontes}.
We explicitly determine the wavefunction of the radion when there is a 
stabilizing mechanism. We find that the radion mass is 
typically $O($TeV$)$. In the limit that the backreaction 
of the stabilizing fields on the metric is small, we find that the 
correction of the stabilizing field to the radion kinetic 
term is subdominant to the gravitational contribution. 
We also find that once the stabilizing field has a non--zero 
vev, the Kaluza-Klein (KK) tower couples directly to the 
brane world fields, with $1/$TeV normalization, and amplitude 
depending on the size of the backreaction. 

The action we consider is \footnote{The action is integrated 
over the circle rather than the line segment.}  
\begin{equation} 
-M^3 \int d^5 x \sqrt{g} {\cal R} + 
\int d^5 x \sqrt{g} 
\left(\frac{1}{2} \nabla \phi \nabla \phi -V(\phi) \right)
- \int d^4x \sqrt{g_4} \lambda_{P}(\phi) 
- \int d^4x \sqrt{g_4} \lambda_{T}(\phi), 
\end{equation} 
where $g_4$ is the induced metric on the branes. 
The background vev for $\phi$ and background metric that  
preserve 4-D Lorentz invariance is 
\begin{equation} 
\phi (x,y)=\phi_0(y) ,
\end{equation} 
\begin{equation}
ds^2= e^{-2A} \eta_{\mu \nu} dx^{\mu} dx^{\nu} -dy^2 . 
\end{equation} 
The Einstein equations are then 
\begin{equation} 
R_{ab}=\kappa^2 \tilde{T}_{ab}=\kappa^2\left(T_{ab}
-\frac{1}{3} g_{ab} g^{cd}T_{cd} \right),  
\end{equation}
with $\kappa^2 = 1/(2 M^3)$. 
For this background the scalar and metric field equations are 
\begin{eqnarray} 
4 A'^2-A''&=& -\frac{2 \kappa^2 }{3}  V(\phi_0) - \frac{\kappa^2 }{3}
\sum_i \lambda_i(\phi_0) \delta(y-y_i) ,
\\
A'^2 &=& \frac{\kappa^2 {\phi_0}'^2}{12} 
- \frac{\kappa^2}{6} V(\phi_0)  
\label{aprimeeqn} \\ 
\phi_0 '' &=& 4 A' \phi_0 ' + \frac{\partial V(\phi_0)}{\partial \phi}+ 
\sum_i \frac{\partial \lambda_i(\phi_0)}{\partial \phi} \delta(y-y_i).
\label{phibackeqn}
\end{eqnarray} 
Here primes denote $\partial/\partial y$, and we reserve $\partial_{\mu}$ to 
denote derivative with respect to the comoving 
4-D spacetime coordinates $x^{\mu}$.
The boundary equations for $A$ and $\phi_0$ are obtained by matching 
the singular terms in the above equations. This gives  
\begin{equation} 
[A'] |_i= \frac{\kappa^2}{3} \lambda_i(\phi_0), 
\label{jump1} 
\end{equation}
\begin{equation}
[\phi_0'] |_i= \frac{\partial \lambda_i(\phi_0)}{\partial \phi} .
\label{jump2}
\end{equation}

For analytical solutions we use an approach presented 
in \cite{dewolfe,otherscalar}. A particular class of potentials 
$V$ is considered which can be written in the form 
\begin{equation} 
V(\phi) = \frac{1}{8} 
\left(\frac{\partial W(\phi)}{\partial \phi}\right)^2
-\frac{\kappa^2}{6} W(\phi)^2 . 
\end{equation}
Then a solution to the following first order equations 
\begin{equation} 
\phi_0 ' = \frac{1}{2} \frac{\partial W}{\partial \phi} \hbox{ , } 
A'=\frac{\kappa^2}{6} W(\phi_0) 
\end{equation} 
automatically solves both the Einstein and scalar field 
equations, once the appropriate boundary conditions are 
solved. The virtue of this method is that for 
simple choices of $W$ it is possible to also solve for the 
backreaction of $\phi_0$ on the metric. This will 
be important for us, since we find that only 
after including the backreaction of the stabilizing 
field does one find that the radion acquires a mass. 

In particular, to obtain some analytic results
the following superpotential \cite{dewolfe} 
will be used:  
\begin{equation} 
W(\phi)= \frac{6 k}{\kappa^2}-  \vv \phi^2
\end{equation} 
with brane potentials 
\begin{equation} 
\lambda(\phi)_\pm = \pm W(\phi_\pm) \pm W'(\phi_\pm)(\phi-\phi_\pm) 
+ \gamma_\pm^2 (\phi-\phi_\pm)^2 . 
\end{equation} 
Here $+/-$ refer to Planck/TeV brane. 
The solution is \cite{dewolfe} 
\begin{eqnarray} 
\phi_0(y) &=& \phi_P e^{-\vv y } \hbox{ , } \\
A(y) &=& ky + \frac{\kappa^2 \phi^2_P}{12} e^{-2 \vv y} . 
\label{Abackgr}
\end{eqnarray} 
The separation distance $r_0$ is then fixed by matching 
$\phi_0$ at $0$ and $r_0$ to $\phi_P$ and $\phi_T$ which 
gives $\vv r_0 = \ln \phi_P/\phi_T$. So the 
quantity 
\begin{equation} 
e^{-\vv r_0} = \frac{\phi_T}{\phi_P}
\end{equation} 
is not a (hierarchically) small number, since both 
$\phi_P$ and $\phi_T$ are $O(M^{3/2} _{Pl})$. 
This combination will appear later in the expression for 
the radion mass. Also for future reference,  
since the backreaction corresponds to the second term in $A$, 
the limit of a small backreaction is 
$\kappa^2 \phi^2_P,\kappa^2 \phi^2_T \ll 1$, and $\vv>0$, but with 
$\phi_P / \phi_T=$constant, so that $\vv$ is kept constant.

\section{The Coupled Field Equations}
\setcounter{equation}{0}
\setcounter{footnote}{0}
\label{coupled-sec}

When $\phi_0=0$ there is always a static solution 
independent of the value of $r_0$.\footnote{After two finetunings 
which are independent of $r_0$. But only one finetune remains after 
radius stabilization \cite{gw,CGRT}. } 
The small fluctuations in the relative position between the two branes
then describe a massless particle (``the radion''), 
and its wavefunction is \cite{CGR} 
$G(x,y)=2 F(x,y)=2 k e^{2ky} R(x)$ and where $\Yfund R=0$.
Since the coupling of the radion to the Standard Model fields 
is $\sim 1/$TeV \cite{CGRT,GW3},  
obtaining an acceptable phenomenology requires that this 
radion acquires a mass.

We therefore consider the spectrum of perturbations about the 
above background which stabilizes the inter--brane separation.  
A general ansatz to describe 
the spin--0 fluctuations is
\begin{equation}
\phi(x,y)=\phi_0(y)+\varphi(x,y) 
\label{a1}
\end{equation}
\begin{equation}
ds^2= e^{-2A-2 F(x,y)} \eta_{\mu \nu} dx^{\mu} dx^{\nu} -(1+G(x,y))^2 dy^2. 
\label{a2}
\end{equation}
In order to describe all gravitational excitations of the model, one
would need to add also the degrees of freedom in the graviton, by
replacing $\eta_{\mu \nu} \to \eta_{\mu \nu}+h^{TT}_{\mu \nu}$. 
One can show that the Einstein equations
with this replacement will have the radion and the graviton decoupled. 
This metric ansatz (\ref{a2}) (together with the two equations 
(\ref{varphieq}) and $G=2F$ which we will shortly derive) 
fixes our gauge choice. One can show, that 
the effect of the remaining gauge transformations that preserve 
the form of  (\ref{a2}) and (\ref{varphieq}) just amount to a 4D gauge
transformation on the graviton field $h_{\mu\nu}$ and can be used to 
impose a convenient 4D gauge for the graviton.
In the following we will only concentrate on the
radion field.
Using this ansatz the Einstein and 
scalar field equations are linearized to obtain some  
coupled equations for $F$, $G$ and $\varphi$. 
The linearized Einstein equations are 
\begin{equation} 
\delta R_{ab}=\kappa^2 \delta \tilde{T}_{ab} .
\end{equation}
Inspecting 
the $\delta R_{\mu \nu}$ equation one immediately concludes 
that $G=2F$. For 
\begin{equation} 
\delta R_{\mu \nu}= \cdots +
2 \partial_{\mu} \partial_{\nu} F- \partial_{\mu} \partial_{\nu} G
+\cdots 
\end{equation}
where the ellipses all contain terms $\sim \eta_{\mu \nu}$. Since 
to linear order in the perturbations 
the sources $\delta \tilde{T}_{\mu \nu}$ are also all  $\sim \eta_{\mu \nu}$, 
the $\partial_{\mu} \partial_{\nu}$ term in $\delta R_{\mu \nu}$ term 
must vanish. This gives $G=2F+ c$. However in the limit 
$F \rightarrow 0$, or $G \rightarrow 0$
we should recover the background solution, so $c=0$.
In what follows we set $G=2F$. Then the coupled field equations are  
\begin{eqnarray}
\delta R_{\mu \nu}&=&\eta_{\mu \nu} \Yfund  F +  
e^{-2A} \eta_{\mu \nu}\left(-F''+10 A' F' + 6A''F-24 {A'}
 ^2 F \right), \\ 
\delta R_{\mu 5} &=& 3 \partial _{\mu} F' - 6 A' \partial _{\mu} F ,
\\
\delta R_{55} &=& 2 e^{2A} \Yfund F + 4 F'' -16 A' F'.
\end{eqnarray}
The source terms are 
\begin{eqnarray} 
\delta \tilde{T}_{\mu \nu} &=& -\frac{2}{3} e^{-2A} \eta_{\mu \nu} 
\left( V'(\phi_0) \varphi -2  V(\phi_0) F \right) \nonumber \\
& & -\frac{1}{3} e^{-2A} \eta_{\mu \nu}
\sum_i \left(\frac{\partial \lambda_i(\phi_0)}{\partial \phi} \varphi
- 4 \lambda_i(\phi_0)F \right) \delta(y-y_i) ,
\\
\delta \tilde{T}_{\mu 5} &=& \phi_0' \partial _{\mu} \varphi, 
\\
\delta \tilde{T}_{55} &=& 2 \phi_0' \varphi' 
+ \frac{2}{3}V'(\phi_0) \varphi + \frac{8}{3} V(\phi_0) F  \nonumber \\
& & +\frac{4}{3} 
\sum_i\left(\frac{\partial \lambda_i(\phi_0)}{\partial \phi} \varphi
+ 2 \lambda_i(\phi_0) F \right) \delta(y-y_i).
\end{eqnarray}
The linearized scalar field equation is 
\begin{eqnarray} 
e^{2A} \Yfund \varphi - \varphi '' + 4 A' \varphi ' + 
\frac{\partial ^2 V}{\partial \phi^2}(\phi_0) \varphi
&=& 
-\sum_i \left(\frac{\partial^2 \lambda_i(\phi_0)}{\partial \phi^2} \varphi
+2  \frac{\partial \lambda_i(\phi_0)}{\partial \phi} F \right)\delta(y-y_i)
\nonumber \\
& & - 6 \phi_0 ' F' -4 \frac{\partial V}{\partial \phi} F .
\end{eqnarray}
Notice that the $R_{\mu 5}$ may be integrated immediately 
to obtain 
\begin{equation} 
\phi_0 ' \varphi = \frac{3}{\kappa^2} (F' -2A' F) . 
\label{varphieq}
\end{equation} 
An integration constant $k(y)$ has been sent to zero since we 
require that the fluctuations $F$ and $\varphi$ are 
also localized in $x$. This equation (\ref{varphieq}) together with the
metric ansatz (\ref{a2}) fixes our gauge choice. One can show, that 
the effect of the remaining gauge transformations that preserve 
the form of  (\ref{a2}) and (\ref{varphieq}) just amount to a 4D gauge
transformation on the graviton field $h_{\mu\nu}$ and can be used to 
impose a convenient 4D gauge for the graviton.

These equations must be supplemented by the boundary conditions 
for $F$ and $\varphi$ 
on the two branes. These are obtained by identifying the singular 
terms in above equations. {\em A priori} the Einstein equations give
two boundary conditions for each wall. It is however straightforward to 
show that one of them is trivially satisfied once $A$ 
satisfies the jump equation (\ref{jump1}). 
The two remaining boundary equations are
\begin{equation} 
[F']= \frac{2 \kappa^2}{3} \lambda_i (\phi_0)  F 
+ \frac{\kappa^2}{3} \frac{\partial \lambda_i}{\partial \phi}(\phi_0) 
\varphi .
\end{equation} 
\begin{equation} 
[\varphi '] |_i =  
\frac{\partial ^2 \lambda_i }{\partial \phi ^2}(\phi_0) \varphi + 
2 \frac{\partial \lambda_i}{\partial \phi}F
\label{phibound}
\end{equation} 
Upon using the jump equations for the background the 
first equation is seen to be equivalent to (\ref{varphieq}) and 
so provides no new constraints. Then only the second boundary 
condition 
must be implemented. A convenient limit will at times be considered 
in this paper. The second boundary condition simplifies in the 
limit of a stiff boundary potential. Namely, if $\partial ^2 
\lambda_i / \partial \phi^2 \gg 1$ then the second boundary condition 
is just $\varphi |_i =0$. Then in this limit the first boundary 
condition is just 
\begin{equation}
(F' -2A' F  ) |_i=0.
\label{bc1}
\end{equation} 

A single equation for $F$ is obtained as follows. One considers 
the combination $e^{2A} \delta R_{\mu \nu}+ \delta R_{55}$ 
{\em in the bulk}. 
The point of this combination is to eliminate terms of 
the form $V(\phi_0) \varphi$. This leaves a bulk equation for 
$F$ and  $\varphi'$ only: 
\begin{equation} 
e^{2A} \Yfund F + F''-2 A' F'= \frac{2 \kappa^2}{3} \phi_0 ' \varphi' 
\label{eHeqn}
\end{equation} 
One then eliminates $\varphi'$ in favor 
of $F$ using (\ref{varphieq}). 
This gives 
\begin{equation} 
F''-2A'F'-4 A'' F -2 \frac{\phi_0''}{\phi_0'} F' 
+4 A' \frac{\phi_0''}{\phi_0'} F  
= e^{2A} \Yfund F , 
\label{Heqn} 
\end{equation} 
to be solved in the bulk. 
This is the principle equation that will be studied and solved 
below. We note in passing that
each eigenmode $\Yfund F_n=-m^2_n F_n$ to this equation  
has two integration constants and one mass eigenvalue. 
One constant corresponds 
to the overall normalization. The remaining integration constant 
is fixed by the boundary condition at the Planck brane, and the 
mass is determined by the boundary condition on the TeV brane. 
In the stiff potential approximation we use the boundary 
condition given by (\ref{bc1}). 

It is then possible to show that a solution $F$ to the above 
equation automatically implies 
that the $\varphi$ equation and the remaining 
Einstein equation are satisfied. In particular, 
starting with (\ref{eHeqn}), one uses the derivative 
of (\ref{varphieq}) to eliminate $F''$. The resulting 
equation, call it $E$, is then differentiated and the 
combination $0=E'-2A'E$ is constructed. Using the background 
field equations and (\ref{varphieq}) one arrives at 
the $\varphi$ equation. Finally, the $\delta R_{\mu \nu}$ 
equation is obtained from the $\delta R_{55}$ equation after 
substituting for $\varphi'$.

\section{General Properties of the Equation}
\setcounter{equation}{0}
\setcounter{footnote}{0}
\label{general-properties-sec}

First we show that the  single ordinary differential equation for 
$F(y)$ given in (\ref{Heqn})
can always be brought into the Schr\"odinger form. For this we
first transform the equations into the coordinate system where the 
background metric is conformally flat. This is achieved by the 
change of variables $dz e^{-A(z)}=dy$, where $A(z)=A(y(z))$. 
In these coordinates the equation simplifies to
\begin{equation}
F''- 3 A'F'-4 A'' F-2\frac{\phi_0''}{\phi_0'} F' +4 \frac{\phi_0''}{\phi_0'} 
A' F=-m^2 F.
\end{equation}
After the rescaling of the field $F$ by $F=e^{3/2 A} \phi_0' \tilde{F}$
we obtain the Schr\"odinger-like equation
\begin{equation}
\label{Schrodinger}
-\tilde{F}''+\left( \frac{9}{4} A'^2 +\frac{5}{2} A'' -
A'\frac{\phi_0''}{\phi_0'} 
+2 \left(\frac{\phi_0''}{\phi_0'}\right)^2 
- \frac{\phi_0'''}{\phi_0'}\right)\tilde{F}
=m^2 \tilde{F}.
\end{equation}

However, this by itself does not guarantee hermiticity of the differential
operator in (\ref{Schrodinger}). The reason is that this operator is defined
only on a finite strip, and therefore in addition to writing the equation in
a Schr\"odinger form one also has to ensure that one has hermitian
boundary conditions for $F$.  For the differential operator in 
(\ref{Schrodinger}) to actually be hermitian on the  finite
strip between the two branes, one also has to require that for any two
functions $F_1,F_2$ on the strip $F_1'(0)F_2(0)-F_1(0)F_2'(0)
-F_1'(z_b)F_2(z_b)+F_1(z_b)F_2'(z_b)=0$,
where $0$ and $z_b$ denote the positions of the branes in the conformally 
flat $z$ coordinates. Once this condition
is satisfied, it is automatically guaranteed by the usual theorems that
all eigenvalues $m_n^2$ are real, that 
the eigenfunctions are orthogonal to each other and that they form a 
complete set. The actual boundary conditions that $F$ has to satisfy 
can be derived from the general boundary condition given in 
(\ref{phibound}). In the particular model considered in this paper the boundary
condition in the $y$ coordinates is given by
\begin{equation}
\pm \varphi'=\gamma_{\pm} ^2 \varphi \pm 2 \vv \phi_{\pm} F
\end{equation}
In the special limit when $\gamma_{\pm} \to \infty$ this boundary condition
reduces to $\varphi=0$ on the two boundaries, 
which together with the constraint equation
(\ref{varphieq}) between $\varphi$ and $F$ just implies 
\begin{equation}
\label{match1}
(F'-2 A' F)|_i=0
\end{equation}
at the two branes. Upon transforming to the
Schr\"odinger basis and $z$ coordinates the boundary condition will be 
replaced by
\begin{equation}
\label{match2}
\tilde{F}'=\tilde{F} (\frac{1}{2} A' -\frac{\phi_0''}{\phi_0'})
\end{equation}
at the branes. This boundary condition clearly satisfies the hermiticity 
properties and thus will ensure the appearance of only real mass eigenvalues
of the coupled system. This will also be the case that we will
analyze in full detail in the following sections. 
As for the general case, when $\gamma_i$ is finite, the boundary condition
will not be hermitian. This can be easily seen from the fact that 
the general boundary condition involves $\varphi'$ at the branes, which
should be 
expressed from (\ref{varphieq}) in terms of $F''$, $F'$ and $F$ at the brane.
The appearance of $F''$ in the boundary condition will generically
ruin the hermiticity of the operator. 
Nevertheless, one may eliminate $F^{''}$ in favor of the eigenvalue, 
and one can in principle solve for $F$. The non--hermiticity 
by itself however does not mean
that the eigenvalues are not real. In fact, since $\phi$ is a real
scalar, and $F$ a component of the metric tensor, both of these functions
have to be real to start with, which guarantees at least the appearance
of only real eigenvalues. While for the model studied here (see 
Section \ref{mass-sec}) the radion is not tachyonic, it is unclear whether  
for a general potential this remains true.
However, the orthogonality of the solutions is not
guaranteed by anything, and will likely be 
violated in general for the non-hermitian
boundary conditions.   It would be interesting to understand the 
physics behind the non-orthogonality of these solutions in more detail.

\section{Approximate solution for the KK tower}
\setcounter{equation}{0}
\setcounter{footnote}{0}
\label{approx-KK-sec}

We have seen that the coupled radion-scalar system leads to a single
ordinary second order differential equation. From now on we will 
always assume that we can use the limit $\gamma_i \to \infty$, and 
be able to use the hermitian boundary conditions (\ref{match1}).
In the following we will present an
approximate solution to these equations. For this, we will first neglect
the backreaction of the non-vanishing scalar background on the metric. This 
will lead us to a simple Bessel-type equation, which will give a very
good approximation for the masses of the KK tower of the fields. However,
surprisingly, in this approximation the radion (which we identify as
the lowest lying solution of (\ref{Heqn}) remains massless. Therefore, after
presenting this approximation, we will give a perturbative analysis for 
the effect of the backreaction of the metric on the radion mass. We will
find that as expected, the radion mass will be of order TeV, but somewhat 
lighter just as predicted in \cite{CGRT,GW3}. 

\begin{figure}
\PSbox{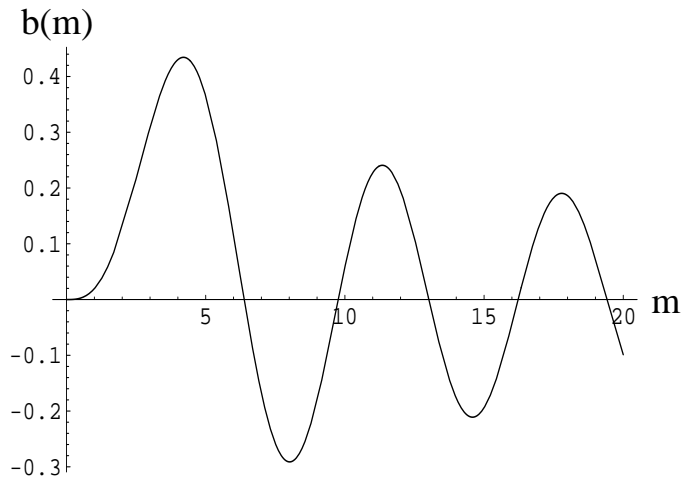 hscale=60 vscale=60 hoffset=150  voffset=0}{8cm}{7cm}
\caption{The lowest mass eigenvalues for the coupled radion-scalar 
system for $\vv/k=1$ are given by the zeroes of the function $b(m)$ 
defined in (\protect\ref{bm}). On this plot $m$ is given in units
$k e^{-k r_0}$, therefore the mass spacings are given by the 
TeV scale. Note, that in the approximation leading to this 
equation the lowest lying state is still massless.}
\label{fig:oscillating}
\end{figure}

\begin{figure}
\PSbox{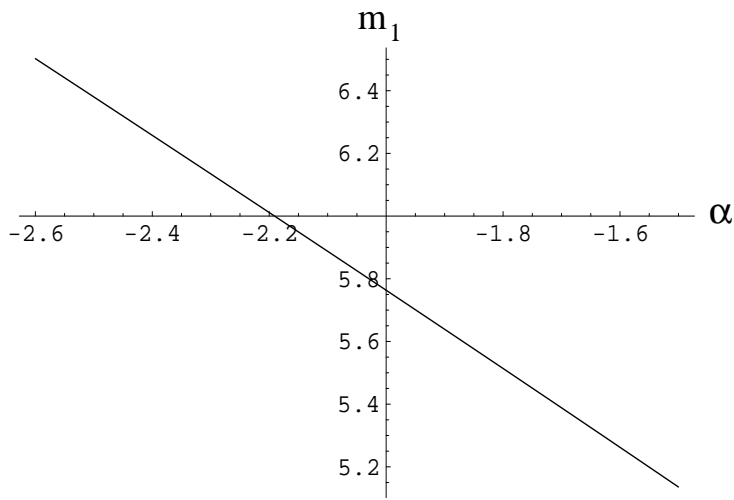 hscale=60 vscale=60 hoffset=150  voffset=0}{8cm}{7cm}
\caption{The dependence of the mass of the first KK mode on $\alpha$. Here
$m_1$ is again given in units $k e^{-k r_0}$ and is therefore of the order
of the TeV scale.}
\label{fig:mass}
\end{figure}

To find the actual wave functions and masses for the radion-scalar 
system, we will use the particular model put forward by de Wolfe et al. 
\cite{dewolfe} and
summarized in Section \ref{review-RS-GW-sec}.
First we neglect the backreaction of the scalar
field background on the metric, which seems to be a good approximation
as long as $\kappa \phi_{P,T} \ll 1$. In this case the equation for the
radion field $F$ reduces in the Schr\"odinger frame to:
\begin{equation}
-F''+\frac{\alpha (\alpha +1) k^2}{(kz+1)^2}F= m^2 F,
\end{equation}
where $\alpha$ is given by
\begin{equation}
\alpha=-\frac{3}{2}-\frac{\vv}{k}.
\end{equation}
In these coordinates the boundary conditions at the brane simplify to
\begin{equation}
\label{match3}
F'+\frac{\alpha k}{kz +1} F=0
\end{equation}
at the locations of the branes at $z=0$ and $z_b \equiv \frac{1}{k} 
(e^{kr_0}-1)$.
The solutions of these equations are given by linear combinations 
of the Bessel functions $J_{\alpha+\frac{1}{2}}$ and Neumann functions
$N_{\alpha+\frac{1}{2}}$:
\begin{equation}
F_n (z)= a_n (z+\frac{1}{k})^{\frac{1}{2}} N_{\alpha+\frac{1}{2}} 
(m_n (z+\frac{1}{k})) +
b_n (z+\frac{1}{k})^{\frac{1}{2}} J_{\alpha+\frac{1}{2}} 
(m_n (z+\frac{1}{k})).
\end{equation}
The mass eigenvalues $m_n$ can be determined from the boundary
condition (\ref{match3}). Using the relation for Bessel functions
\begin{equation}
Z_n'(x)=Z_{n-1} (x)-\frac{n}{x} Z_n (x)
\end{equation}
the boundary conditions at the two branes simply reduce to
\begin{eqnarray}
&& a_n N_{\alpha-\frac{1}{2}} (\frac{m_n}{k})+b_n J_{\alpha-\frac{1}{2}}
(\frac{m_n}{k})=0, \nonumber \\
&& a_n N_{\alpha-\frac{1}{2}} (\frac{m_n e^{k r_0}}{k})+
b_n J_{\alpha-\frac{1}{2}}
(\frac{m_n e^{kr_0}}{k})=0, 
\end{eqnarray}
which yields the simple equation
\begin{equation}
b(m_n)=J_{\alpha-\frac{1}{2}}
(\frac{m_n}{k}) \frac{N_{\alpha-\frac{1}{2}} (\frac{m_n e^{k r_0}}{k})}{
N_{\alpha-\frac{1}{2}} (\frac{m_n}{k})}-J_{\alpha-\frac{1}{2}}
(\frac{m_n e^{kr_0}}{k})=0,
\label{bm}
\end{equation}
which can be used to determine the mass eigenvalues $m_n$. This can be
done numerically. In Fig.~\ref{fig:oscillating} we show the lowest 
mass eigenvalues for $\alpha = -2.5$, which corresponds to 
the somewhat unrealistic value $\frac{\vv}{k}=
1$. In Fig.~\ref{fig:mass} we show the dependence of the first non-vanishing
mass eigenvalue on the value of $\alpha =-3/2-\vv/k$.
One can easily see from (\ref{bm}), that $m=0$ is always a solution to 
(\ref{bm}), therefore in the approximation we are using the radion is
still massless. For the higher states of the KK tower 
it is a good approximation to use the mass eigenvalues obtained from
(\ref{bm}), because the masses are of the order (and even larger) than the
TeV scale, thus in the limit of small backreaction that we are considering 
throughout the paper these masses will be only slightly modified. 
The radion (which appeared as the zero mode above) however 
needs special treatment, since the shift in the mass (which is usually 
negligible for the higher KK modes) coming from the backreaction
of the metric background due to the scalar field is the leading 
order contribution to the 
mass for the radion. Below we will estimate the size of the
radion mass in perturbation theory.

\section{The radion mass}
\setcounter{equation}{0}
\setcounter{footnote}{0}
\label{mass-sec}

In the previous section we have seen what the approximate wave-functions and 
masses are for the KK tower of the coupled radion-scalar system. 
In this approximation of neglecting the backreaction, 
however, we have still found a vanishing radion mass. This is 
in fact easy to show for a general stabilizing potential. 
From (\ref{Heqn}), $F= e^{2 A}$ is always a solution with zero mass
if $A''$ is neglected in the bulk. Thus the radion mass is always 
proportional to the backreaction of the metric independently of the
details of the potential of the stabilizing scalar field.
In the following,
we will show how the backreaction generates a non-vanishing mass
for the radion field. For this, we start with the equation 
describing the radion wave function in the $y$ coordinates:
\begin{equation}
F''-2 A' F' -4A''F +2 \vv F' -4 \vv A' F+m^2 e^{2A} F=0,
\end{equation}
where $A(y)$ is given in (\ref{Abackgr}). The appropriate boundary condition
is $F'-2A' F=0$ at the 
branes. In the special limit $\gamma_{\pm} \rightarrow \infty$ the other 
boundary condition is $\varphi=0$. Thus
we will treat the 
backreaction as a perturbation, and look for the solution in terms
of a perturbative series in $l \equiv \kappa \phi_P /\sqrt{2}$. 
Then we write the solution as 
\begin{equation}
F_0=
e^{2 k |y|} (1+l^2 f_0(y)) \hbox{ , }m_r^2= l^2 \tilde{m}^2 \hbox{ , } 
A(y)= k |y| +
\frac{l^2}{6} e^{-2 \vv |y|}.
\end{equation}
Expanding the solution as above and keeping only the leading terms in $l^2$
we obtain the equation
\begin{equation}
f_0''+2 (k+\vv) f_0'=-\tilde{m}^2 e^{2 k |y|} -\frac{4}{3} 
(k-\vv)\vv e^{-2 \vv |y|}
\end{equation}
along with the boundary conditions
\begin{equation}
f_0'+\frac{2}{3} \vv e^{-2 \vv |y|} =0
\end{equation}
at the location of the branes. 
One can easily find the most general solution for $f$ from the equation
in the bulk, which is given by
\begin{equation}
f_0'(y)= C e^{-2 (k+\vv) |y|} -\frac{\tilde{m}^2}{2(2k+\vv)} e^{2 k |y|}
-\frac{2 (k-\vv)\vv}{3k}
e^{-2 \vv |y|}
\label{radwave},
\end{equation}
where the integration constant $C$ along with the radion mass $\tilde{m}$ is
determined by the boundary conditions at the brane. This way we obtain the
radion mass to be
\begin{equation}
m^2_{radion}= \frac{4  l^2 (2k+\vv) \vv^2}{3 k} e^{-2(\vv+k) r_0},
\end{equation}
where $r_0$ denotes the location of the brane. 
Note that this result is very similar to the answer obtained from the
effective theory 
computation using the na\"{\i}ve ansatz\cite{CGRT,GW3}, except for the 
important difference in the power of $\vv/k$.
The 
exact result obtained here scales as $(\vv/k)^2$, whereas
the effective theory result would scale as 
$(\vv/k)^{3/2}$.\footnote{We thank Jim Cline for these
observations.} It would be very interesting to understand the origin of this
different scaling. 
For this model to give
the correct value of the weak scale without reintroducing a large fine-tuning
again one needs $\vv /k \approx {1/37}$,
thus the radion mass turns out to be somewhat lighter than the TeV scale.
It is suppressed by the factor $ l \frac{\vv}{k} e^{-\vv r_0}$ compared
to the TeV scale. Thus in this approximation $m_{radion} \sim \frac{l}{40}$ 
TeV, which could be at least in the range of a few GeV's. Of course, we 
need to emphasize that $l$ is not necessarily small for the stabilization
mechanism to work, we took this limit only for calculational convenience.

\section{Coupling to SM fields}
\setcounter{equation}{0}
\setcounter{footnote}{0}
\label{coupling-KK-sec}
In this section the coupling of the radion and KK tower 
of $\phi$ to the TeV brane are obtained. In particular we
demonstrate that the bulk scalar field gives a small 
correction to the radion kinetic term, and thus the 
kinetic terms obtained from the Einstein--Hilbert part of the action 
dominate, justifying the results obtained using the na\"{\i}ve 
ansatz \cite{CGRT,GW3}. 

In the previous section it was seen that by including 
the backreaction an ${\cal O}($TeV$^2)$ for the radion is obtained.
The wavefunction is then 
\begin{equation} 
F_0(x,y)=e^{2k|y|}\left(1 + l^2 f_0(y)\right)R(x)
\end{equation} 
where $f_0(y)$ is given by the integral of (\ref{radwave}). 
Since the radion mass 
is ${\cal O}($TeV$^2)$, and by assumption $l^2 \ll 1$, 
we see by inspection that the backreaction induces 
a small correction to the unperturbed wavefunction. So for the 
purposes of determining the coupling of the radion to the 
TeV brane it is sufficient to include only the unperturbed 
wavefunction, namely $F(x,y)=e^{2k|y|} R(x)$. 
Then a straightforward calculation gives 
\begin{equation} 
-M^3 \int dy \sqrt{g} {\cal R} 
\supset 6 M^3 (\partial R)^2 \int e^{-2A} e^{4 k |y|} 
= \frac{6 M^3}{k}  (e^{2k r_0} -1) (\partial R)^2 
\label{radnorm1}
\end{equation} 
So the normalized radion $r(x)$ 
is $R(x) = r(x) e^{-k r_0}/\sqrt{6} M_{Pl}$, 
since $M^3/k = M^2_{Pl}/2$.  
This implies a coupling to the TeV brane fields which is 
\begin{equation} 
R(x) e^{2k r_0} \left(1+ {\cal O}(l^2)\right)\hbox{Tr}T_{\mu \nu} 
= \frac{r(x)}{\sqrt{6} M_{Pl} e^{-kr_0}} \hbox{Tr}T_{\mu \nu}
\left(1+ {\cal O}(l^2)\right),
\end{equation} 
where the left hand side of this equation is a consequence of the fact that
the induced metric on the TeV brane is $e^{-2A(r_0)}(1-e^{2kr_0}R(x) 
\eta_{\mu\nu})$. The coupling obtained this way agrees precisely with
(\ref{rcoupling}). This is perhaps surprising, since 
the latter computation used an {\em ansatz} which did not 
satisfy the equations of motion. This makes us suspect that results
which depend only on the leading unperturbed form of the radion
wave function will be correctly captured by the na\"{\i}ve ansatz.

Now we address the issue that was originally raised by \cite{Cline}. 
Is the radion kinetic term dominated by the  
kinetic term of the bulk scalar field or 
the bulk gravity action, and in particular is the 
former hierarchically larger?  
To answer this, we need the change in 
$\varphi$ caused by a fluctuation in the radion. Since $\varphi=0$ when the 
backreaction is not included, we 
must include the leading backreaction correction to the 
radion wavefunction, 
given by the integral of (\ref{radwave}).  
From (\ref{varphieq}) we compute that the change in $\varphi$ to 
${\cal O}(l^2)$
due to the radion is 
\begin{equation} 
\varphi = \frac{3}{\kappa^2 \phi_0'} ( F' -2A' F)  
= \frac{3 l^2}{\kappa^2 \phi_0'}(F_0' +\frac{2 \vv}{3}e^{-2 \vv |y|}) 
 = \frac{3 l^2 R(x)}{\kappa^2 \phi_0'} e^{2k |y|}
f_3(y) 
\end{equation} 
where 
\begin{eqnarray}
f_3(y) &\equiv & f'_0(y)+\frac{2}{3}\vv e^{-2 \vv |y|} \nonumber \\
& =&
C e^{-2(k+\vv)|y|}-\frac{\tilde{m}^2}{2(2k+\vv)}e^{2k|y|}+\frac{2 \vv^2}{3k} 
e^{-2\vv |y|} .
\end{eqnarray}
This fluctuation in $\phi$ then contributes 
to the radion kinetic term at $O(l^2)$ an amount 
\begin{equation} 
\int dy e^{- 4 A} g^{\mu \nu} \partial _\mu  \varphi \partial _\nu  \varphi
 =\frac{9 l^2}{2 \kappa^2 \vv^2} 
(\partial R)^2 \int dy e^{2(k-\vv) |y|} f_3(y)^2 .
\label{phikt} 
\end{equation}
From (\ref{radnorm1}) the unnormalized contribution from bulk 
gravity to the kinetic term is 
$\sim e^{2kr_0}$. 
So we only need to consider those contributions from $\phi$ which 
are comparable or larger to this. Recalling that $m^2 \sim e^{-2k r_0}$, 
it is seen that the largest terms in (\ref{phikt}) are at best 
$e^{2 k r_0}$. Explicitly performing the integral one finds that 
it is 
\begin{equation} 
\delta {\cal L}= 2 l^2 \frac{ \vv^2 }{\kappa^2  k^2} 
(\partial R)^2
 e^{2 k r_0 -6 \vv r_0} \left(\frac{1}{3k-\vv} + 
\frac{1}{k-3\vv} - \frac{1}{k-\vv} \right)\left(1+ O(l^2)\right) . 
\end{equation} 
This is typically $\sim l^2 \vv^2 e^{2k r_0} M^3 /k^3 $, 
which is smaller than (\ref{radnorm1}) since 
we assuming that the backreaction is small, $l \ll 1$, and also 
that $\vv \ll k$ to obtain a realistic hierarchy. So  
the radion kinetic term is dominated by the contribution 
from the bulk gravity, and receives a small correction from 
the stabilizing bulk scalar field. 

In Section \ref{general-properties-sec} it was found that for the 
simple boundary conditions $\varphi =0$ (corresponding 
to the limit $\partial ^2 \lambda_{\pm} / \partial \phi^2  \gg 1$) 
a self--adjoint equation for $F$ was obtained. The general 
solution to this is 
\begin{equation} 
F(x,y) =\sum _n  \alpha_n F_n(x,y) 
\label{Htower}
\end{equation} 
where $F_n$ is a mass eigenstate, and the 
$\alpha_n's$ are some numbers. We expect that $F$ includes the massive 
radion, but where did all the other states come from? It is 
helpful to reconsider what happens when the backreaction is 
neglected. 
In this limit the KK tower in $F$ completely disappears and only the 
(massless) radion remains. 
This may be observed from (\ref{varphieq}), 
since neglecting the backreaction corresponds to $\kappa^2 \phi_0 \ll 1$, 
and this amounts to setting $F'-2A'F=0$. The only solution for $F$ in this 
case is the radion zero mode $F=e^{2k|y|}$. Once the backreaction is 
included, however, the fluctuating modes in $\phi$ and $F$ are 
correlated through (\ref{varphieq}). In particular, a general 
fluctuation 
$\varphi$ induces a change in $F$. The sum over KK states 
appearing above is then just the decomposition of $F$ 
into these KK eigenstates. 
It is then expected that the coefficients 
$\alpha_n$ for the non--radion states to be suppressed by 
the backreaction.  

The preceding remarks imply that the TeV brane fields, 
which couple  
to the induced metric $F$, 
also directly couple to the KK tower,
by an amount suppressed by the 
backreaction.\footnote{The KK modes of the scalar field do not mix with the
KK modes of the graviton. The reason is that the only way they could mix is by
a coupling of the 5D trace of the metric to the scalar KK modes. However, the
graviton is traceless, and the trace is basically identified with the radion,
therefore no additional graviton-scalar mixing could be introduced.}
Since $F \sim \phi_0' \varphi$ is already suppressed 
by the backreaction, therefore in order 
to compute the induced metric to lowest 
order in the backreaction we can use the zeroth order 
wavefunctions for $\varphi$. 
The normalized KK fields are given by 
\begin{equation} 
\varphi_n(x,z)= 
 \frac{\psi_n(x)}{N_n}(kz+1)^2  J_{2+\vv/k}(m_nz+1/k) 
\left(1+ l^2 f_n(y) \right). 
\end{equation} 
Here 
$\psi_n(x)$ are the normalized 4--D fields satisfying 
$\Yfund \psi_n=-m^2_n \psi_n$.
The orthogonality of these solutions when the backreaction 
vanishes $(l=0)$ follows 
from the boundary condition $\varphi_n=0$ and 
the properties of the Bessel functions. Also
$\psi_n(x)$ are the normalized 4--D fields satisfying 
$\Yfund \psi_n=-m^2_n \psi_n$. The normalization constant is 
\begin{equation} 
N_n= \frac{1 }{ \sqrt{k}} e^{kr_0} 
J_{3+\vv/k} \left(\frac{m_n}{k} e^{kr_0}\right) . 
\end{equation} 
As discussed previously, 
the lowest order masses $m_n$ are determined by 
$J_{2+\vv/k}(e^{2k r_0}m_n/k)=0$ and are real since the operator 
equation with these boundary conditions 
is self-adjoint. 
The coupling of these fields to the TeV brane is given by 
\begin{equation} 
F(x,y=r_0) \hbox{Tr}T_{\mu \nu} 
\end{equation} 
where $F$ is the solution of (\ref{varphieq}) for the 
solutions $\varphi_n$ given above. One finds the coupling 
\begin{equation} 
 \frac{l }{m_n} \lambda_s x^{\vv} _n
\psi_n(x) \hbox{Tr}T_{\mu \nu}.
\end{equation}
The model-dependent couplings that appear are
\begin{equation} 
\lambda_S = \frac{\sqrt{2}}{3} \vv \kappa \sqrt{k} e^{-\vv r_0}  
 \sim {\cal O}(\frac{\vv}{k}) 
\end{equation} 
and 
\begin{equation} 
x^{\vv} _n = \frac{J_{1+\vv/k}(\frac{m_n}{k} e^{kr_0})}
{J_{3+\vv/k}(\frac{m_n}{k} e^{kr_0})} 
\end{equation} 
is a numerical constant of $O(1)$. While the inclusion of 
the backreaction leads to a TeV suppressed coupling for the 
KK modes, the size decreases rather rapidly due to the 
$1/m_n \sim 1/$TeV suppression, as may be observed from 
inspecting Fig. 1. 

The coupling discussed here implies that the KK modes of 
$\phi$ can be directly produced at future colliders, and 
they also decay directly to Standard Model fields. 
This may be puzzling at first, 
since the stabilizing potential may have a global discrete symmetry, 
such as $Z_2$, which would 
na\"{\i}vely imply that some of these KK modes are stable. 
The  
background vev for $\phi$ explicitly breaks this symmetry, however, and 
this allows for all the KK modes to decay into the brane world 
fields. 

The direct coupling of the KK modes from the 
stabilizing fields may 
have interesting implications for search strategies and 
current limits on the Randall-Sundrum framework. In particular, 
it may be important to {\em not} neglect the stabilizing potential 
when discussing these 
issues.  However, when the backreaction is small, 
the size of their couplings is suppressed 
by $\vv/k \sim 1/40$ compared 
to the that of the radion. Therefore, in what follows, we neglect 
these states in the loop computations.  
 
\section{Cosmological Implications}
\setcounter{equation}{0}
\setcounter{footnote}{0}
\label{cosmology-sec}

The subject of brane cosmology has recently attracted lots of interest
\cite{BDL,CGKT,CGRT,Minnesota,othercosm,perturbations,othercosm2,GRS,millenium,otherother}.
Most of this was due to the realization, that the expansion of a brane
universe could be significantly different from the ordinary 
Fried\-mann--Ro\-bert\-son--Wal\-ker 
(FRW) cosmology \cite{BDL,CGKT}. However, it did not
take very long to realize, that this is simply due to the fact, that a 
generic brane model (like the one presented in \cite{BDL}) can not
give the ordinary cosmological evolution, since gravity is in general
manifestly higher dimensional. This means that in these models
the 4D effective theory is usually
not described by 
ordinary Einstein gravity, but generically a complicated scalar-tensor
theory of gravity. However, observations show that our Universe is 
described by Einstein's theory of relativity to a high precision, therefore
one has to require from the outset that a brane model reproduces ordinary
Einstein gravity, at least at long enough distances. Once this is achieved,
the cosmological expansion will be automatically described by the 
ordinary FRW Universe, which simply follows from the fact that the 
effective theory is ordinary Einstein gravity. Thus one can see that 
the issue of unconventional cosmologies is nothing else but the issue of
whether one recovers 4D gravity. This issue manifests itself in the case
of the Randall-Sundrum two-brane 
model due to the presence of the radion field.
Without a stabilizing potential, the radion field will be massless, and 
yield additional long range forces, and also contribute to the expansion of the
Universe, yielding an unconventional cosmology, which is presumably excluded
by the requirement for a successful nucleosynthesis \cite{CGKT}.
Thus the radion field has to obtain a mass. Once it is massive, gravity 
on both branes will be ordinary 4D gravity, and thus the cosmology will
be conventional below temperatures comparable to the radion mass. This
has been explained in great detail in \cite{CGRT}, and also in
\cite{Minnesota}. In \cite{CGRT} a simplified calculation for the
cosmological expansion has 
been presented, where the wave function of the radion has been neglected,
and also the effects of the stabilizing scalar field were included
by adding a five dimensional potential for the radion field $V(b)$.
Assuming that that the potential $V(b)$ is very steep, it was shown 
from a perturbative solution of the bulk equations that the ordinary
FRW Universe is recovered. It was also argued, that the $55$ component 
of the Einstein equation, which in the absence of a stabilizing potential
usually leads to the unconventional expansion equations will only
determine the shift in the radion field due to matter on the wall, and does
not result in unconventional cosmologies once the radius is stabilized.
Below we demonstrate, that the results of \cite{CGRT} which were neglecting
the radion wave function, and also did not include the fluctuations of the
scalar field at the brane remain valid in the more precise framework
of radion stabilization explained in the previous sections. In particular,
we will show, that the result obtained for the shift in the radion field due 
to matter on the walls in Eq. (4.15) of \cite{CGRT} is exactly reproduced
in the full calculation.

To compute $G_{55}$ we use the ansatz 
\begin{equation} 
ds^2 = n(t,y)^2 dt^2- a^2(t,y) d^2 x - b(t,y)^2 dy^2 
\end{equation}
for which 
\begin{equation}
G_{55}= 3 \left[ \frac{a'}{a}\left(\frac{a'}{a} + \frac{n'}{n}\right) 
-\frac{b^2}{n^2} \left(\frac{\dot{a}}{a}
\left(\frac{\dot{a}}{a}-\frac{\dot{n}}{n}\right) 
-\frac{\ddot{a}}{a} \right)\right] .
\end{equation}
The jump equations for $a$ and $n$ on the TeV brane imply \cite{BDL}
\begin{equation} 
\frac{[a']}{a} = - \frac{\kappa^2}{3}(\lambda_{-}(\phi)+ \rho) b 
\hbox{ , } \frac{[n']}{n} = \frac{\kappa^2}{3}(-\lambda_{-}(\phi)+ 3p+2 \rho)b 
\end{equation} 
Here $\rho$ and $p$ are the bare energy matter density and pressure 
on the TeV brane, which are related to the physically measured 
quantities on the TeV brane 
by $\rho_{0} = \rho e^{-4 A_0}$, etc. where $e^{-A_0}$ is the 
scale factor on the TeV brane. 
Then averaging the $G_{55}$ equation about the TeV brane and 
linearizing to ${\cal O}(\rho,F,\varphi)$ gives 
\begin{eqnarray} 
<G_{55}> &=& \kappa^4 \frac{\lambda^2 _{-}(\phi_0)}{6}  
-3 e^{2A}\left( \left(\frac{\dot{a}}{a} \right)^2+ \frac{\ddot{a}}{a}
\right)
- \frac{\kappa^4 \lambda_{-}(\phi_0)}{12}(3p-\rho) 
\nonumber \\
& &
+\frac{\kappa^4 \lambda_{-}(\phi_0)}{3}\frac{\partial \lambda_{-}(\phi_0)}{\partial 
\phi} \varphi + \frac{2\kappa^4 \lambda^2 _{-}(\phi_0)}{3} F.  
\end{eqnarray}
Terms with $\dot{n}$ are higher order in $\rho$ and are dropped, and 
$b=1 +2F$ has been used. 
For late--time cosmology in the presence of radion stabilization
it is reasonable to use the FRW equation
\begin{equation} 
\left(\frac{\dot{a_0}}{a_0} \right)^2
+ \frac{\ddot{a_0}}{a_0}=-\frac{1}{6 M^2_{Pl}} 
\left(3 p_* - \rho_* + 3 p_0 - \rho_0 \right) .
\end{equation} 
Note this includes a contribution from matter $(\rho_*)$ on the Planck 
brane. Also implicit in the use of this equation is 
the assumption that 
the time variation of the radion is negligible,
which is justified {\em a posteriori}. Then    
using the relation $\kappa^4 \lambda_{-}(\phi_0)=-6(1-e^{-2A_0})/M^2_{Pl}$
gives 
\begin{eqnarray} 
<G_{55}>&=& \frac{\kappa^4 \lambda^2 _{-} (\phi_0)}{6} 
+ \frac{e^{4A_0}}{2 M^2_{Pl}}\left(3 p_0-\rho_0 + e^{-2A_0} 
(3 p_* - \rho_*) \right) \nonumber \\
& & +
\frac{\kappa^4 \lambda_{-}(\phi_0)}{3}\frac{\partial \lambda_{-}(\phi_0)}{\partial 
\phi} \varphi + \frac{2\kappa^4 \lambda^2 _{-}(\phi_0)}{3} F.  
\end{eqnarray} 
The $G_{55}$ equation is 
\begin{equation} 
G_{55} = \kappa^2 T_{55}=\kappa^2 
\left( \frac{1}{2}{\phi'} ^2 - g_{55}(\frac{1}{2}(\nabla \phi)^2 -V(\phi))
\right). 
\end{equation} 
Then the averaging of $T_{55}$ and linearizing using 
(\ref{a2}) gives 
\begin{equation} 
\kappa^2<T_{55}>=\kappa^2(\frac{1}{2}{\phi'} ^2 _0 - V(\phi_0) ) 
+ \kappa^2(\phi' _0 \varphi' - 4 F V - \frac{\partial V}{\partial \phi}
\varphi) 
\end{equation} 
with all quantities are evaluated on the TeV brane. Using the 
background bulk equation (\ref{aprimeeqn}) and the jump 
equation (\ref{jump1}) the leading terms are seen to cancel. 
Then after using the background equations, the jump equations for 
the background fields, 
and some algebra gives
\begin{equation}    
\frac{e^{4A_0}}{2 M^2_{Pl}} \left(3 p_0-\rho_0 + e^{-2A_0} 
(3 p_{*} - \rho_{*}) \right) = \kappa ^2
({\phi '}_0 \varphi ' 
-2 {\phi ^{'}}^2 _0 F
-\frac{\partial V}{\partial \phi} \varphi -4 A' \phi'_0 \varphi).
\end{equation} 
Using (\ref{eHeqn}) to eliminate $\varphi'$, (\ref{Heqn}) 
to eliminate $F''$ in favor of the mass eigenvalue, 
and (\ref{phibackeqn}) to eliminate 
$\phi_0 ''$ finally gives 
\begin{equation} 
\frac{e^{4A_0}}{2 M^2_{Pl}}\left(3 p_0-\rho_0 + e^{-2A_0} 
(3 p_* - \rho_*)\right)  =-3 e^{2A_0} m^2_{r} F . 
\end{equation} 
But the shift in the distance between the two branes is obtained 
by integrating the line element, which gives $\delta r_0 = R(e^{2k r_0}-1)/k$.
Then since $F=R e^{2k r_0}$, one obtains
\begin{equation}
\frac{\delta r_0}{r_0}=  \frac{1}{6 k r_0} \frac{(1-e^{-2 A_0})}
{ m^2_r M^2_{Pl} e^{-2 A_0}}
\left(\rho_0- 3 p_0+ e^{-2A_0} 
(\rho_*-3 p_*) \right)
\end{equation} 
which is {\em precisely} the result found in \cite{CGRT} 
obtained by using a 4D effective theory.\footnote{A translation 
dictionary between two different 
 notations is required: $k r_0=m_0 b_0/2$.} This is perhaps 
not surprising, since for constant radion field the 
na\"{\i}ve ansatz and the full metric including the 
wavefunction of the radion are equivalent up to a coordinate transformation.
So in an adiabatic approximation the leading order result 
using the na\"{\i}ve ansatz should agree with that obtained 
from using the correct radion wavefunction, if the fluctuations in the
scalar field are ignored. It is less clear why the full answer including
the contribution from the scalar field turns out to be exactly 
equal to the calculation using the na\"{\i}ve ansatz.   
Note that matter on the Planck brane causes a smaller shift in the 
radion compared to an equal amount of matter on the TeV brane. 
This is because the radion wavefunction is peaked at the 
TeV brane, and it couples more weakly to the 
Planck brane relative to the TeV brane by precisely 
the amount $e^{-2A_0}$. Thus one finds the very general result 
that in the presence 
of matter on the branes and a stabilizing mechanism, 
the $G_{55}$ equation determines the shift in the radion.
 
\section{The Effective 4D Lagrangian}
\setcounter{equation}{0}
\setcounter{footnote}{0}
\label{effective-lag-sec}
In the previous sections we have argued that in the 
presence of a stabilizing potential the linear couplings 
of the radion and bulk scalars is given by 

\begin{eqnarray}
\frac{1}{2} (\partial r)^2 - \frac{1}{2} m^2 r  ^2 + 
\sum_n \frac{1}{2} \left((\partial \psi_n)^2 - m^2_n \psi _n  ^2 \right)
  +DH ^{\dagger} D H \nonumber \\
+\left(\frac{r(x)}{\sqrt{6} \Lambda} + \sum_n \alpha_n \frac{\psi_n(x)}{\Lambda_n} 
\right) \hbox{Tr} T_{\mu \nu}  + \xi H^{\dagger} H {\cal R} -V(H) . 
\end{eqnarray} 
The masses appearing here are $O($TeV$)$, and their particular 
value depends on the details of the stabilizing mechanism. The scale 
$\Lambda= e^{-k r_0} M_{Pl} $ in the Randall-Sundrum model, 
but here we have left it general. The other scales are $\Lambda_n \sim m_n$, 
and the $\alpha_n$ are also model--dependent, and vanish in the limit 
of small backreaction. In the remaining sections we restrict 
ourselves to the above Lagrangian, and do not commit ourselves to 
any specific mechanism of radius stabilization. For the 
electroweak analysis we neglect the contributions of the 
KK modes from $\phi$.

Note that in the above Lagrangian we have also 
included a curvature Higgs scalar operator $H^{\dagger} H {\cal R}$. 
The presence of this operator leads to interesting signals 
for discovering the Higgs and radion at future colliders \cite{GRW}. 
In particular, the branching fractions of the Higgs and radion 
to $gg$ and $\bar{b} b$ can be substantial different from that 
of the SM Higgs.

As discussed in \cite{GRW}, the presence of the conformal term 
$H ^{\dagger} H R$ 
leads to both kinetic and mass mixing between the neutral Higgs and 
radion. Below we summarize the relevant formulae for mixing and 
couplings. The interested reader is referred to the next section 
for details.  

One finds that the ``gauge'' $h$ and $r$ are related to 
the mass eigenstates $h_m$ and $r_m$ by 

\begin{equation} 
h = (\cos \theta - \frac{6 \xi \gamma}{Z} \sin \theta ) h_m 
+ (\sin \theta + \frac{6 \xi \gamma}{Z} \cos \theta) r_m \hbox{ , }
\end{equation} 
\begin{equation} 
r=\cos \theta \frac{r_m}{Z} - \sin \theta \frac{h_m}{Z} \hbox{ ,} 
\end{equation} 
where 
\begin{equation} 
\tan 2 \theta = 12 \xi \gamma Z 
\frac{m^2 _h}{m^2_r-m^2_h - 6 \xi \gamma^2 (1 -12 \xi)} \hbox{ , } 
\end{equation} 
\begin{equation} 
\gamma = \frac{v}{\sqrt{6} \Lambda} \hbox{ , } Z^2=1+ 6 \xi \gamma(1-6 \xi) ,
\end{equation} 
where $v \approx 246$ GeV is the electroweak vev.
Requiring that the quantity $Z^2$ be positive (in order to avoid ghost-like
states) places an upper bound on the value of $\xi$, for a given $\gamma$.
Physically this requirement comes from maintaining positive definite 
kinetic terms for $h$ and $\phi$.

In this basis, the couplings of the physical radion and Higgs 
appropriate for tree--level studies are 
\begin{equation} 
-\left((\cos \theta - (6 \xi -1) \frac{\gamma \sin \theta}{Z}) h_m 
+(\sin \theta + (6 \xi -1) \frac{\gamma \cos \theta}{Z}) r_m \right) 
\hbox{Tr} T_{\mu \nu} . 
\end{equation}
In the $\xi \rightarrow 0$ limit one recovers 
\begin{equation} 
-\left(h- \gamma r \right) \hbox{Tr} T_{\mu \nu} \hbox{ , }
\label{simplecoup}
\end{equation} 
obtained in 
\cite{CGRT,GW3}.  Note that $\hbox{Tr} T_{\mu \nu}$ includes
SM Higgs contributions.

\section{Curvature--Scalar Mixing}
\setcounter{equation}{0}
\setcounter{footnote}{0}
\label{curvature-scalar-sec}
In this Section the effect of introducing 
a curvature--scalar interaction are reviewed. 
The discussion parallels \cite{GRW}, however some of the resulting 
formulae are slightly different because here terms 
of ${\cal O} (\gamma^2)$ and ${\cal O}(\gamma^2 \xi^2)$ 
are kept.

We begin with the couplings of the radion and Higgs to 
the SM fields before electroweak symmetry breaking. 
The induced metric on the TeV wall is 
\begin{equation} 
g^{ind} _{\mu \nu} (x)=e^{-2 A(r_0)-2 e^{2k r_0}R(x)} g_{\mu \nu}(x) , 
\end{equation} 
where the warp factor includes the backreaction, although its 
inclusion is not necessary for our purposes. The 
canonically normalized radion $r$ is 
\begin{equation} 
R(x) = e^{-k r_0} \frac{r(x)}{\sqrt{6} M_{Pl}}. 
\end{equation}
So we express the induced metric expanded about a Minkowski metric as 
\begin{equation} 
g^{ind} _{\mu \nu}(x) =e^{-2 A(r_0)
-2 \frac{\gamma}{v} r(x)} \eta_{\mu \nu}
\equiv e^{-2 A(r_0)} \Omega ^2(r)  \eta_{\mu \nu} 
\end{equation} 
with 
\begin{equation} 
\gamma = \frac{v}{\sqrt{6} \Lambda} \hbox{ , } 
\Lambda = M_{Pl} e^{-k r_0} .
\end{equation} 
The four dimensional effective action we consider is  
\begin{eqnarray} 
S_{TeV} &=& \int d^4 x \sqrt{g_{ind}} \left(g^{\mu \nu}_{ind} 
D_{\mu} H^{\dagger} D_{\nu} H 
-V(H) \right) \nonumber \\
& & + \int d^4 x \sqrt{g}\frac{1}{2} \left( 
(\nabla r)^2 - m^2_{r} r^2 \right) 
+ \int d^4 x \sqrt{g_{ind}} \xi R(g_{ind}) H^{\dagger} H + S_{SM}.
\end{eqnarray} 
To canonically normalize the Higgs and other SM fields, we perform 
the field--independent redefinition 
\begin{equation} 
H \rightarrow e^{A(r_0)} H \hbox{ , } \psi \rightarrow e^{3 A(r_0)/2} \psi 
\end{equation} 
In this basis the Higgs--radion potential is 
\begin{equation} 
V(H,r) = \Omega ^4(r) V(H) . 
\end{equation} 
Note that $V$ also includes the effective 4--dimensional cosmological  
constant, which we assume to vanish. Clearly this potential 
has a minimum at the same location as $V(H)$, so that the EWSB 
vacuum is $r=0$ and $H^0 = v/\sqrt{2}$. 

We consider the  presence of the curvature mixing term  
\begin{equation} 
{\cal L_{\xi} } =\sqrt{g_{ind}} \xi {\cal R}(g_{ind}) H^{\dagger} H. 
\end{equation}  
Our choice of signs for $\xi$ is such that the Higgs potential 
receives a positive mass--squared correction in a de Sitter phase 
when $\xi$ is positive. 
Since this is a renormalizable interaction, there is no reason 
for it not to be present, or to be suppressed. 
What makes this operator important in this 
case is that ${\cal R}$ contains the induced metric, rather than just the 
ordinary 4-dimensional metric. In particular, 
\begin{equation} 
{\cal R}(\Omega^2(r) \eta_{\mu \nu})=-6 \Omega^{-2} \left(\Yfund \ln \Omega 
+ (\nabla \ln \Omega)^2 \right).  
\end{equation} 
So the curvature--scalar interaction is 
\begin{equation} 
{\cal L_{\xi}} = -6 \xi \Omega^2 \left(\Yfund \ln \Omega 
+ (\nabla \ln \Omega)^2 \right)H^{\dagger} H 
\end{equation} 
To see the effect of the curvature scalar interaction we 
expand $H^0=(v+h)/\sqrt{2}$ and $\Omega(r)=1-\gamma r/v +\cdots$. 
We need to only expand $\Omega$ to linear order since the derivative 
terms are already of $O(r)$. This gives  
at quadratic order 
\begin{equation} 
{\cal L_{\xi}} = 6 \xi \gamma h \Yfund r 
+ 3 \xi \gamma ^2 (\partial r)^2 . 
\end{equation} 
where a total derivative has been dropped. 
The $\xi$ terms clearly introduce kinetic mixing. The full radion--Higgs 
Lagrangian to be diagonalized is 
\begin{equation} 
{\cal L} = -\frac{1}{2} h \Yfund h   - \frac{1}{2} m^2_h h^2 
 -\frac{1}{2} (1+ 6 \xi \gamma^2) r \Yfund  r
 - \frac{1}{2} m^2_{r} r^2
+ 6 \xi \gamma h \Yfund r. 
\end{equation} 
The mass parameters $m_{r}$, $m_h$ are the masses of the radion 
and Higgs, respectively, in the limit $\xi =0$. 
The kinetic terms are diagonalized by the shift \
$ h=h' + 6 \xi \gamma r'/Z$  , and $r = r' /Z$. Here 
\begin{equation}
Z^2= 1+ 6 \xi \gamma^2(1-6 \xi)
\end{equation}
is the coefficient of the radion kinetic term after undoing the kinetic mixing,
and is therefore required to be positive 
in order to keep the radion kinetic term positive definite. 
For a  fixed cutoff $\Lambda$ this restricts the size of the mixing parameter 
$\xi$. It must lie in the range
\begin{equation}
\frac{1}{12} (1-\sqrt{1+\frac{4}{\gamma^2}}) \leq \xi \leq 
\frac{1}{12} (1+
\sqrt{1+\frac{4}{\gamma^2}})
\end{equation} for non-zero values of $\gamma$. Otherwise one has a 
ghost-like radion field, which presumably signals an instability of the
theory.

This rescaling diagonalizes the kinetic terms, but introduces 
mixing in the mass matrix. A final rotation $h'=\cos \theta h_m + 
\sin \theta r_m$ and $r'= \cos \theta r_m - \sin \theta h_m$
brings the Lagrangian to canonical form. With the above definition of 
the sign of the rotation, the rotation angle is 
\begin{equation} 
\tan 2 \theta = 12 \xi \gamma Z 
\frac{m^2 _h}{m^2_{r}-m^2_h (Z^2 -36 \xi ^2 \gamma^2)}.  
\label{tan2t}
\end{equation} 
We note that for moderate values of $\xi $ and $\gamma$ (i.e., 
$Z^2 > 36 \xi ^2 \gamma^2$) 
the mixing angle $\tan 2 \theta$ is negative when $m_h > m_{r}$.
For small $\gamma$ we can expand
\begin{equation} 
\tan 2 \theta = 12 \xi \gamma \frac{m^2_h}{m^2_{r} - m^2_h} +O(\gamma^2) .
\label{gangle}
\end{equation} 
Putting everything together, the relation between the gauge 
and mass eigenstates is 
\begin{eqnarray} 
h &=& (\cos \theta - \frac{6 \xi \gamma}{Z} \sin \theta ) h_m 
+ (\sin \theta + \frac{6 \xi \gamma}{Z} \cos \theta) r_m \hbox{ , }
\label{basisrel1} \\
r &=& \cos \theta \frac{r_m}{Z} - \sin \theta \frac{h_m}{Z} . 
\label{basisrel2}
\end{eqnarray} 
The mass eigenvalues are easily obtained 
\begin{equation} 
m^2_{\pm}=\frac{1}{2 Z^2}\left(m^2_{r}+(1+ 6 \xi \gamma^2) m^2_h 
\pm ((m^2_{r}-m^2_h(1+ 6 \xi \gamma^2 ))^2+ 
144 \gamma^2 \xi^2 m^2_{r} m^2_h )^{1/2} \right) 
\label{pmmass}
\end{equation}
The heavier state $(+)$ is identified with the state with the larger of 
$(m^2_h,m^2_{r})$.

\section{Radion Couplings and Feynman Rules}
\setcounter{equation}{0}
\setcounter{footnote}{0}
\label{feynman-sec}
In this section we derive the Feynman rules relevant 
to the computation of the oblique parameters $S$, $T$, 
$U$. 

Before proceeding, we pause to ask whether higher-order 
couplings such as 
\begin{equation} 
\phi^2 \hbox{Tr} T_{\mu \nu} 
\end{equation} 
also affect in particular the electroweak precision measurements  
$S$, $T$ and $U$. This operator could either be directly present, 
or generated from the above linear coupling due 
to a non-trivial kinetic term for the radion. Although this 
operator contributes at one-loop to the gauge boson two point functions, 
it is easy to see that they do not contribute to $T$, since 
the $m^2_V$ contained in $ \hbox{Tr} T_{\mu \nu}$ is canceled 
by the $1/m^2_V$ appearing in the expression for $T$, nor to $S$ or $U$ 
since the contribution of this operator to the vacuum 
polarizations is momentum--independent.  
Thus we need to only consider the linear coupling 
\begin{equation} 
\frac{\gamma r}{v}\hbox{Tr} T. 
\end{equation}   

This operator will have a contribution to the oblique corrections similar
to that of the standard model Higgs. 
First we discuss the Feynman rules for the interactions from the
$\frac{\gamma r}{v} {\rm Tr} T$ operator. 
The interaction Lagrangian
term relevant for the gauge-boson propagator corrections is just given by
\begin{equation}
{\cal L}_{int} = -\frac{\gamma}{v} r (2 M_W^2 W_\mu^+ W^{\mu -} 
+M_Z^2 Z_\mu Z^\mu ).
\end{equation}
In addition, to ensure gauge invariance of the results, one also has to
examine the gauge fixing terms carefully. The gauge fixing Lagrangian for
the $W$ and the $Z$ are given by
\begin{equation}
{\cal L}_{gf} = \sqrt{g}
\left[-\frac{1}{\alpha} (-D_\mu W^{\mu +}+i \alpha M_W \Psi^+)(-D_\mu W^{\mu -}
-i \alpha M_W \Psi^-)-\frac{1}{2\alpha} (-D_\mu Z^{\mu} +\alpha M_Z \Psi)^2,
\right]
\end{equation}
where the $\Psi$'s are the would-be-Goldstone bosons, and $\alpha$ is the
gauge fixing parameter in the R$_\alpha$ gauge. Note, that since the 
gravitational
background is non-trivial, we have replaced the ordinary derivatives by
covariant derivatives. The background metric is given by 
$g_{\mu\nu}= \Omega^2(r ) \eta_{\mu\nu}=e^{-\frac{2\gamma r}{v}} 
\eta_{\mu\nu}$, therefore the covariant derivative of a vector will take the
form
\begin{equation}
D_\mu V^\mu = \Omega^{-2} (\partial_\mu V^\mu -\frac{2\gamma}{v} \partial_\mu
r V^\mu).
\end{equation}
Thus, from the gauge fixing terms one also obtains three-point interaction 
vertices of the form
\begin{equation}
{\cal L}_{gf}= \frac{2\gamma}{v \alpha} (\partial_\mu Z^\mu) (\partial_\nu r 
Z^\nu )+\frac{2\gamma}{v \alpha} (\partial_\mu W^{\mu +}) (\partial_\nu r 
W^{\nu -})+\frac{2\gamma}{v \alpha} (\partial_\mu W^{\mu -}) (\partial_\nu r 
W^{\nu +}).
\end{equation}
With these operators added, the Feynman rule for the three-point function
is given by

\begin{minipage}{4cm}
\PSbox{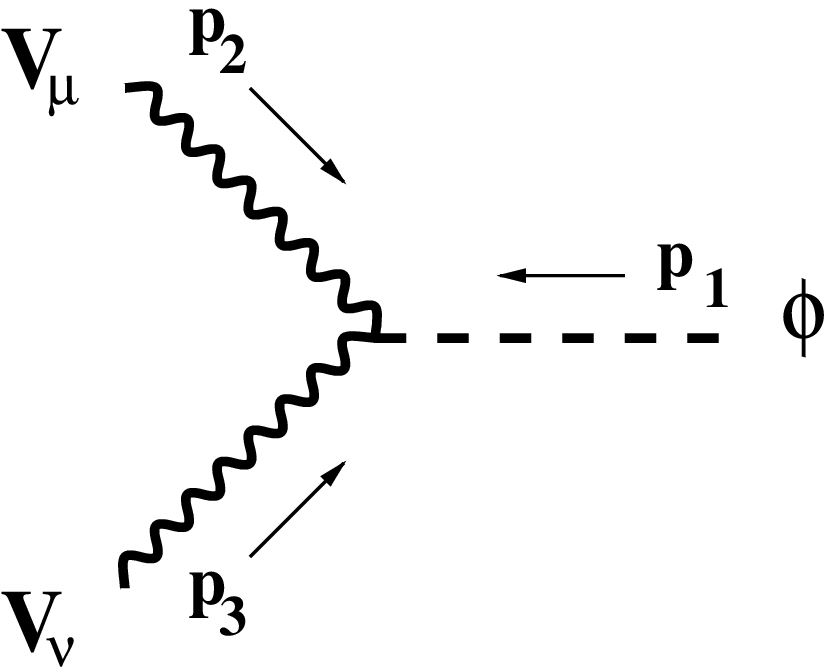 hscale=40 vscale=40 
hoffset=40  voffset=20}{3cm}{4cm}
\end{minipage}
\begin{minipage}{11.8cm}
\begin{equation}
\frac{-i2M_V^2 \gamma}{v} \eta_{\mu \nu} 
-\frac{i2\gamma}{\alpha v} (p_{2\mu} 
p_{1\nu}+p_{3\nu}p_{1\mu})  
\label{Feyn1}
\end{equation}
\end{minipage}

In addition to the cubic vertices evaluated above, there are also
4-point couplings of the radion and the gauge bosons. These terms will not
contribute to the oblique electroweak corrections, however they will be
important to obtain a gauge invariant answer for the gauge boson
vacuum polarization diagrams. These terms arise
from two different sources. The first source is the conformal coupling
$-e^{-\frac{\gamma r}{v}} {\rm Tr} T$ to the trace of the energy momentum
tensor. In the formalism of \cite{CGRT} this can be obtained by a very
careful expansion of the interaction terms $ |D_\mu H|^2+
\frac{1}{\sqrt{6}\Lambda} \partial r (H^\dagger DH +h.c.)$. Either way
one finds the additional operators 
\begin{equation}
\frac{\gamma^2}{v^2} r^2 (\frac{M_W^2}{2} W_\mu^+W^{\mu -}+
M_Z^2 Z_\mu Z^{\mu}).
\end{equation}
The other source of quartic interaction terms are again the gauge fixing
terms. One simply expands these to higher order to obtain the 
interaction terms
\begin{equation}
-\frac{2\gamma^2}{\alpha v^2} \partial_\mu r \partial_\nu r (
Z^\mu Z^\nu+2 W^{\mu +}W^{\nu -}).
\end{equation}
The interaction terms involving the would-be Goldstone bosons just 
contribute a total derivative, and thus they can be omitted.
The above operators give rise to the following Feynman rule for the
4 point function:

\begin{minipage}{4cm}
\PSbox{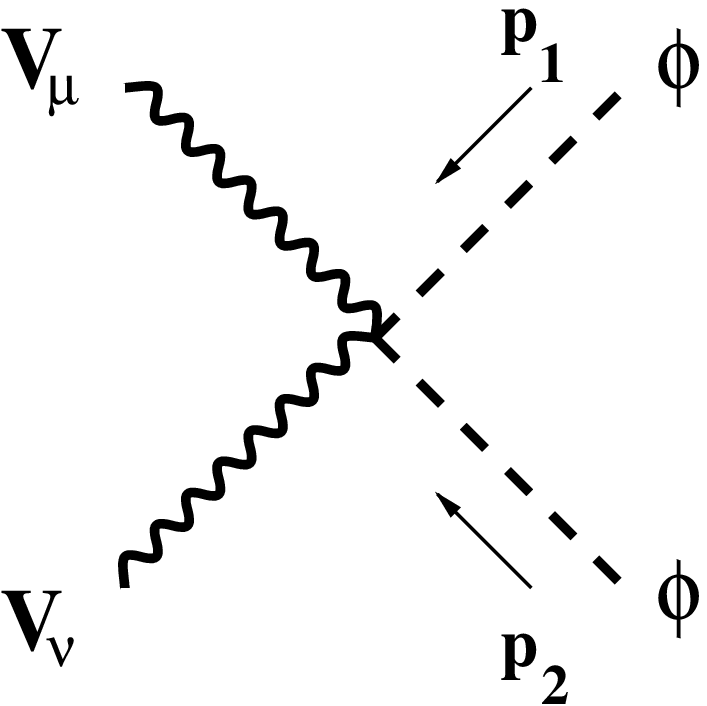 hscale=40 vscale=40 
hoffset=40  voffset=20}{3cm}{4cm}
\end{minipage}
\begin{minipage}{11.8cm}
\begin{equation}
\frac{i2M_V^2 \gamma^2}{v^2} \eta_{\mu \nu} 
+\frac{i4\gamma^2}{\alpha v^2} (p_{1\mu} 
p_{2\nu}+p_{1\nu}p_{2\mu})  
\label{Feyn2}
\end{equation}
\end{minipage}

The presence of the terms proportional to $\frac{1}{\alpha}$ in (\ref{Feyn1})
and (\ref{Feyn2}) are in fact crucially important to obtain 
gauge invariant amplitudes, however for calculations in the unitary gauge 
$\alpha \to \infty$ their effect vanishes.

This is however not the complete story for the Feynman rules.  
The reason is that the  radion couples conformally 
to the metric and so is not the same as the 
Higgs. Thus for example 
one finds that at one--loop the radion has the anomalous 
coupling 
\begin{equation} 
\frac{r}{\Lambda} b_G  \frac{\alpha_G}{8 \pi} G_{\mu \nu} G^{\mu \nu} 
\end{equation} 
in addition to the usual momentum--dependent coupling obtained from  
one-loop diagrams with internal fermions. Here $b_G$ is the beta--function. 
This may be understood as due to the scaling anomaly together with 
$r$ as 
a generator of scale transformations. Diagrammatically this result is 
obtained by preserving the conformal coupling of $r$ when 
the theory is regulated. 
For 
dimensional regularization this means that the radion must couple 
conformally to the $D$-dimensional metric. Since the linear 
coupling of the radion is obtained from varying the induced metric, 
for loop computations the radion should couple instead 
to $\hbox{Tr}_D T_{\mu \nu}$, 
but where now the trace is evaluated in $D$-dimensions. This differs 
from the above coupling by some operators whose coefficient vanishes 
when $D \rightarrow 4$. Since this $\epsilon=4-D$ dependence 
can be offset by poles appearing in the loops, the appearance 
of these additional operators can result in   
finite non-zero results in the $D \rightarrow 4$ limit. These operators will
indeed have a non-vanishing contribution to the S and U parameter.
Next we calculate the Feynman rules for these ``anomalous'' couplings.
The interactions we should therefore study are 
\begin{equation} 
{\cal L}= -\frac{h}{v} \hbox{Tr}_4 T 
+ \frac{ \gamma r}{v} \hbox{Tr}_D T . 
\end{equation} 
We point out that it is 
the original `gauge' radion, $r$, that has the conformal coupling 
to the metric, and consequently it is this field which appears in the 
above interactions. But a result of the 
curvature--scalar term interaction is to introduce  
mixing between the radion and Higgs, which implies that after transforming 
to the mass basis the 
physical Higgs, 
$h_m$, will have couplings similar to those above.

The radion will thus have the interaction terms
$(D=4-\epsilon)$
\begin{equation} 
{\cal L} = \gamma \frac{r}{v}\hbox{Tr}_D T_{\mu \nu} = 
\gamma \frac{r}{v} \hbox{Tr}_4 T -\gamma \frac{\epsilon}{4} 
\frac{r}{v} 
F_{\mu \nu} F^{\mu \nu} + 
\gamma\frac{\epsilon}{2} \frac{r}{v} (2 M^2_W W^+ W^- + M^2_Z Z^2), 
\label{toy}
\end{equation} 
where the first term is the one we have already discussed above.
We will later add in the Higgs and mixing coefficients. 
The terms relevant to the gauge boson propagators from the last two terms are:
\begin{eqnarray}
{\cal L}_{int}^{anom}= &&-\frac{\epsilon \gamma}{2v} r \left[
(\partial_\mu Z_\nu)^2 -(\partial_\mu Z_\nu)(\partial_\nu Z_\mu)+
2 (\partial_\mu W_\nu^+)(\partial^\mu W^{\nu -}) 
-2 (\partial_\mu W_\nu^+)(\partial^\nu W^{\mu -})-\right. \nonumber \\
&& \left. 2 M_W^2 W_\mu^+ W^{\nu -} -M_Z^2 Z_\mu Z^\nu \right].
\end{eqnarray}
In $D$ dimensions, one also has to modify the gauge fixing terms. The 
covariant derivative of a vector field will be modified to
\begin{equation}
D_\mu V^\mu = \Omega^{-2} (\partial_\mu V^\mu -\frac{(D-2)\gamma}{v} 
\partial_\mu r V^\mu).
\end{equation}
In addition, the $\sqrt{g}$ factor in front of the gauge fixing 
terms have to be modified to $\Omega^D$. 
Thus these interaction terms modify the Feynman rules for the interaction 
vertex given in (\ref{Feyn1}) to
\begin{equation}
\frac{-2iM_V^2 \gamma}{v} (1-\frac{\epsilon}{2})\eta_{\mu \nu} 
+\frac{i\gamma \epsilon}{v} (p_2\cdot p_3 \eta_{\mu\nu}-p_{2\nu}p_{3\mu})
-\frac{i(2-\epsilon)\gamma}{\alpha v} (p_{2\mu} 
p_{1\nu}+p_{3\nu}p_{1\mu}) -\frac{i\gamma\epsilon}{\alpha v} 
p_{2\mu}p_{3\nu}.
\end{equation}
In addition, the four-point vertex in (\ref{Feyn2}) is also modified by
terms proportional to $\epsilon$, which however do not contribute to a 
calculation in the unitary gauge.

We close this section by discussing how to take the mixing between
the Higgs and the radion due to the possible presence of the curvature-scalar
mixing operator into account. 
The interaction in the gauge basis is 
\begin{equation} 
{\cal L}= -\frac{h}{v} \hbox{Tr}_4 T 
+ \frac{ \gamma r}{v} \hbox{Tr}_D T . 
\end{equation} 
Using 
\begin{eqnarray} 
h&=& a h_m + b r_m  \hbox{ , } \\ 
r&=& c h_m + d r_m , 
\end{eqnarray} 
where the coefficients $a$, $b$, $c$ and $d$ can be read off 
from (\ref{basisrel1}),(\ref{basisrel2}), then 
\begin{eqnarray} 
{\cal L} &=& \left(-(a - \gamma c)\frac{h_m}{v} 
+ (\gamma d - b ) \frac{r_m}{v} \right)  
\hbox{Tr}_4 T 
\nonumber \\
& & 
+ (c h_m + d r_m) \frac{\gamma}{v}  \left(
-\frac{\epsilon}{4} 
F_{\mu \nu} F^{\mu \nu} + \frac{\epsilon}{2} M^2_V V^2 \right) .
\label{eff-mixed-lag-eq}
\end{eqnarray} 
Thus the Feynman rules for the radion (and also for the Higgs) have to be 
modified such, that the above mixing terms are taken properly into account,
for example for the mass eigenstate radion the Feynman rule will be
\begin{equation}
\frac{-2iM_V^2}{v} (\gamma d-b-\frac{\gamma d \epsilon}{2})\eta_{\mu \nu} 
+\frac{i\gamma d \epsilon}{v} (p_2\cdot p_3 \eta_{\mu\nu}-p_{2\nu}p_{3\mu})
-\frac{i(2-\epsilon)\gamma d }{\alpha v} (p_{2\mu} 
p_{1\nu}+p_{3\nu}p_{1\mu})-\frac{i\gamma\epsilon d}{\alpha v} 
p_{2\mu}p_{3\nu}.
\end{equation}

\section{Electroweak Precision Measurements}
\setcounter{equation}{0}
\setcounter{footnote}{0}
\label{electroweak-sec}
In this Section we consider the corrections of the 
Randall--Sundrum model to the oblique parameters. Our
analysis also applies more generally to a model 
with a light scalar coupled conformally to the SM metric
but with a typical coupling of ${\cal O}($TeV$)^{-1}$.

Our approach is to use an effective theory with cutoff
${\cal O}(\Lambda)$, similar to the approach taken in
\cite{effective}. Below this scale the only light fields are 
the radion and Higgs whose contributions we are going to 
calculate explicitly. In our approach the effect of 
any modes heavier than the cutoff are included by  
introducing higher dimension operators that directly 
contribute to the oblique parameters. 
This in principle 
includes the effects of the heavy spin--2 KK states, for example, 
which are typically heavier than the radion. 
A direct computation of the effect of the heavy spin-2 states 
using a momentum--dependent regulator has been presented in 
\cite{DHR-big}. 
 
In the previous Section the radion 
coupling to the gauge bosons was obtained and found to be similar 
to that of the Higgs. The contribution of the Higgs to 
the oblique parameters is by itself divergent, but these 
divergences are canceled by the contribution from the 
pseudo-Goldstone bosons (or the longitudinal states of the 
massive gauge bosons). Thus for the radion one expects a 
divergent contribution, but in contrast to the Higgs there is no 
additional source to cancel this. This is perhaps not surprising 
since the radion interactions are non-renormalizable. 

A set of operators that provide the necessary counterterms for 
the wavefunction renormalization is 
\begin{equation} 
{\cal O}_{X}= 
\frac{g^2 Z_X}{\Lambda^2} 
\left( H^{\dagger} H \hbox{Tr} W_{\mu \nu} W^{\mu \nu} + 
\frac{1}{2}\tan^2 \theta_W
H^{\dagger} H B _{\mu \nu} B^{\mu \nu}
+ \tan \theta_W H^{\dagger} W_{\mu \nu} B^{\mu \nu} H \right),
\label{op1}
\end{equation} 
where $W_{\mu \nu} = W^{a} \tau^a$, with the generators normalized 
to $1/2$.
Note that the last operator is gauge invariant, since for a gauge 
transformation $U$, 
$W_{\mu \nu} \rightarrow U W_{\mu \nu} U^{\dagger}$. 
Setting the Higgs to its vev in the above operator 
gives 
\begin{equation} 
{\cal O}_{X} \rightarrow 
\frac{g^2 Z_X v^2}{2 \Lambda^2} 
\left( W^{+}_{\mu \nu} {W^{-}} ^{\mu \nu} + \frac{1}{2 \cos^2 \theta_W } 
Z_{\mu \nu} Z^{\mu \nu} \right) .
\label{op12} 
\end{equation} 
In this model the radion does not contribute to $\gamma \gamma$ and 
$\gamma Z$ wavefunction renormalization, and the absence of these 
counterterms 
uniquely fixes the coefficients in 
(\ref{op1}). The explicit computation of the $ZZ$ and $WW$ 
wavefunction renormalizations demonstrates that 
the above operator has the correct relative factor between the 
two gauge bosons. 
This is perhaps non--trivial, 
since there is no additional degree of freedom to 
fix this relative factor. 
We also note 
that an identical operator to (\ref{op12}) 
is also required in technicolor 
theories in order to cancel 
the divergent contribution of pseudo--Goldstone bosons to the oblique 
parameters \cite{randallgolden}.

The operator which provides the counterterms for the mass renormalization is 
\begin{equation} 
{\cal O}_{M} = \frac{Z_M}{2 \Lambda^2} 
\left( {g^{\prime}} ^2 ( D_{\mu} H ^{\dagger} H)
(H^{\dagger} D^{\mu} H) + g^2 H^{\dagger} H ( D_{\mu} H^{\dagger} D^{\mu} H) 
\right) .
\label{op2}
\end{equation}
The first operator that appears here violates the custodial symmetry, 
and in particular contributes to the $Z$ but not the $W$ mass.  
After electroweak symmetry breaking they together 
reduce to 
\begin{equation} 
{\cal O}_{M} \rightarrow \frac{Z_{M}}{\Lambda^2} \left( M^4_W W^{+}_{\mu} 
W^{- \mu} + \frac{M^4 _Z}{2} Z^{\mu} Z_{\mu} \right) .
\end{equation}
Note that it is $m^4_V$ which appears, so we explicitly see that this
operator 
contributes to  
the $\rho$ parameter. 
In this case it is trivial to obtain the correct mass renormalization 
from (\ref{op2}), since here there are two coefficients to be determined 
from only two 
constraints.
So in addition to the usual Standard Model renormalizations, 
these wavefunction and mass counterterms are also required 
to renormalize the model.

The model with the radion represents an effective theory valid for 
$E \lesssim \Lambda$. The dimension--6 operators discussed 
above are obtained by 
integrating some unknown degrees of freedom in the full theory, 
and in the effective 
theory they appear with some unknown coefficients $Z_i(\Lambda)$. 
These for example could include the effects of integrating out 
the heavy spin-2 KK modes. 
Since the divergences for which the two above operators act as counterterms
arise at one loop order, they are proportional
to $\gamma^2 /(16\pi^2)$. Thus it is reasonable to expect that the
finite part of the operator is also of the same order, and in order to match
the form of the explicitly calculated one-loop corrections we will write 
the (finite) coefficients as
\begin{equation} 
Z_M=\frac{\gamma^2 \Lambda^2}{16 \pi^2 v^2} a_M, \ \  
Z_X=\frac{\gamma^2 \Lambda^2}{16 \pi^2 v^2} a_X,
\end{equation}
where we expect that the dimensionless parameters $a_M$ and $a_X$ are at 
most of order one. Note that since $\gamma^2 \sim \Lambda^{-2}$, 
in this parameterization  
the dimension--6 operators are still suppressed by $\Lambda^2$.

These coefficients parameterize the unknown physics integrated out 
at the scale $E \sim \Lambda$. Since, however, the radion at one 
loop contributes to the anomalous dimension of these operators, 
when comparing to the 
experimental results evaluated at the $Z$ mass large logarithms of 
$O(\ln \Lambda/M_Z)$ appear from this anomalous scaling and this 
effect should be included.
Following Wilson, 
a one-loop Wilsonian 
renormalization group equation is obtained for the operator 
coefficients. 
In the leading logarithm approximation 
the value of these coefficients at the weak scale is determined to be 
\begin{equation} 
Z_i(M_Z)= Z_i(\Lambda)+ \frac{\beta_i}{16 \pi^2} \ln \frac{\Lambda^2}{M^2_Z}  
,
\end{equation}
with the $\beta_i$ determined from the coefficient of 
the $\ln \mu$ term (or equivalently, from the $1/ \epsilon$ poles) 
in an explicit one--loop computation. 
To compute the oblique parameters, one then adds the contribution of these 
renormalized operators to the finite parts of the one-loop diagrams. 
In the leading logarithm approximation this amounts to simply replacing 
the $1/\epsilon$ poles in the gauge--boson self-energies with 
$\ln \Lambda / M_Z$.

To compute the oblique parameters one uses the Feynman rules in the 
previous Section to compute the two Feynman diagrams that contribute 
to the vacuum polarizations. Both diagrams have one internal 
radion, and one uses the three point function and the other uses the 
four point function.  
Our convention for the 
sign of the vacuum polarizations $\Pi_{VV}$ is that
\begin{minipage}{4cm}
\PSbox{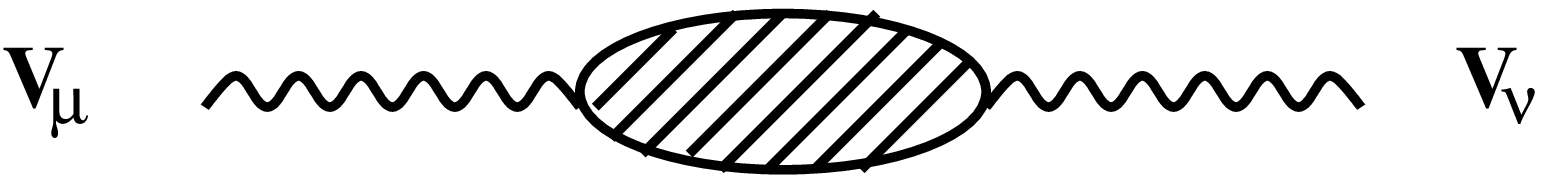 hscale=35 vscale=35 
hoffset=0  voffset=45}{3cm}{4cm}
\end{minipage}
\begin{minipage}{12.35cm}
\begin{equation}
\ \ =i \Pi^{\mu \nu} _{VV}(p^2) = 
i \eta ^{\mu \nu} \Pi_{VV}(p^2) 
+ i p^{\mu} p^{\nu} \tilde{\Pi}_{VV}(p^2)
\end{equation}
\end{minipage}

\noindent and only the first term is computed. 
The generic form of the radion contribution is 
\begin{equation} 
\Pi_{VV}(p^2)= \Pi^S_{VV}(p^2) + \Pi^A_{VV}(p^2) . 
\end{equation} 
Here '$A$' denotes the anomalous contribution due 
to the conformal coupling of the radion, and 
'$S$' denotes the standard contribution which is 
also similar to the Higgs contribution (when 
$\xi=0$). The anomalous couplings are discussed in the 
previous Section. By an appropriate rescaling of the 
coupling we can also use $\Pi^S$ for the Higgs. 
When $\xi \neq0$, the results given below 
can also be used to compute the oblique parameters
after an appropriate redefinition of the couplings and 
masses. The modification to the expression for the oblique parameters is 
summarized in (\ref{generalX}).

Inspecting the Feynman diagrams for the vacuum polarizations
one finds that the quad\-ra\-tic divergences cancel between the 
two diagrams, leaving only a logarithmic divergence. Therefore 
this justifies the use of dimensional regularization. 
An explicit computation of the 
Feynman diagrams in unitary gauge and for vanishing 
curvature scalar parameter $\xi$ gives 
\begin{eqnarray} 
\Pi^S_{VV}(0) &=&- \frac{ \gamma^2}{16 \pi^2} \frac{m^4_V}{v^2}
\left( \frac{6}{\epsilon} +\frac{5}{2} - \frac{m^2_r}{2 m^2_V} 
+3 \frac{m^2_V \ln m^2_V /\mu^2- m^2_r \ln m^2_r / \mu^2}
{m^2_r-m^2_V } \right) \hbox{ , } 
\label{vp1} \\
\Pi^S_{VV}(m^2_V) &=& \frac{ \gamma^2}{16 \pi^2} \frac{m^4_V}{v^2}
\left( -\frac{20}{3 \epsilon} 
-\frac{2}{3} \frac{m^2_r}{m^2_V} + \frac{1}{3} \frac{m^4_r}{m^4_V} 
+\frac{10}{9} \right)  \nonumber \\
& & + \frac{ \gamma^2}{16 \pi^2} \frac{m^4_V}{v^2}
\left(
4- \frac{4}{3} \frac{m^2_r}{m^2_V} 
+ \frac{m^4_r}{m^4_V} \right) \int_0^1 dx \ln(x^2 m^2_V +(1-x) m^2_r )/\mu^2 
\nonumber \\  
& & +\frac{ \gamma^2}{16 \pi^2} \frac{m^4_V}{v^2} 
\left( - \frac{m^2_r}{m^2_V}(\frac{m^2_r}{3 m^2_V}-1) 
\ln \frac{m^2_r}{\mu^2} +\frac{1}{3} (\frac{m^2_r}{m^2_V}-2) 
\ln \frac{m^2_V}{\mu^2}  
\right) \hbox{ , }
\label{vp2}
\end{eqnarray} 
where to avoid confusion with defining 
too many $\gamma's$, the renormalization scale $\mu$ appearing 
includes the usual factors of $4 \pi$ and Euler's constant $\gamma_E$.
An analytic expression for the Feynman parameter integral may be obtained, 
but it is not very illuminating. 
A powerful check on these expressions is gauge invariance. 
We have explicitly checked that in the general R$_\alpha$ 
gauge, the gauge parameter $\alpha$ cancels from the expression and
reproduces the above results.
We note that the divergences appearing here have the form as given 
by the operators in (\ref{op1}) and (\ref{op2}). For $\Pi(0)_{VV} 
\sim m^4_V$ as required by (\ref{op2}). The difference 
between $\Pi(m^2_V)$ and $\Pi(0)$ gives the divergence proportional 
to $p^2$, but in the above equations $p^2$ has 
already been set to $m^2_V$. With this in mind, by inspection 
the coefficient 
of $p^2$ is $\sim m^2_V$, as required by 
(\ref{op1}). So the set of operators given by (\ref{op1}) and 
(\ref{op2}) provide an appropriate set of counterterms. We 
renormalize using $\overline{\hbox{MS}}$.
Then using the Wilsonian approach outlined above, in the 
leading logarithm approximation we just replace 
$1/ \epsilon $ with $\ln \Lambda /M_Z$, that is 
\begin{equation} 
\frac{2}{\epsilon} -\gamma_E + \ln 4 \pi \rightarrow \ln \frac{\Lambda^2} 
{M^2_Z} ,
\end{equation}
and set $\mu= M_Z$.

The anomalous contribution is finite and is 
\begin{equation} 
\Pi^A _{VV}(p^2) = \frac{\gamma^2}{16 \pi^2} \frac{m^4_V}{v^2} 
\left( - \frac{10}{3} \frac{p^2}{m^2_V} +6 -2 \frac{m^2_r}{m^2_V} 
\right) . 
\label{avac}
\end{equation} 
We note that the $m^2_r$ term does not contribute to the 
oblique parameters since it is $p^2$--independent and 
also $\sim m^2_V$. 

The PDG convention for the oblique parameters that we use here is  
\begin{eqnarray} 
T &=&  \frac{1}{\alpha} \left( \frac{\Pi_{WW}(0)}{M^2_W} 
- \frac{\Pi_{ZZ}(0)}{M^2 _Z}  \right)
 \\
S&=& \frac{4 \sin^2 \theta_W \cos ^2 \theta_W}{\alpha}
\left(\frac{\Pi_{ZZ}(M^2_Z)}{M^2_Z}-
\frac{\Pi_{ZZ}(0)}{M^2_Z} \right) \\
S+U&=& \frac{4 \sin^2 \theta_W }{\alpha}
\left(\frac{\Pi_{WW}(M^2_W)}{M^2_W}-
\frac{\Pi_{WW}(0)}{M^2_W} \right)
\end{eqnarray} 
More generally one must also include the $\Pi_{Z \gamma}$ and 
$\Pi_{\gamma \gamma}$ self-energies. They have been 
dropped here since they do not receive contributions from 
either the 
Higgs or the radion.

Using the above expressions one can evaluate the contribution of 
the radion to $S$ and $T$ in the limit of a large radion mass. 
One obtains for $\xi =0$, 
\begin{eqnarray} 
S &=& \frac{\gamma^2}{\pi} \left( a_X-\frac{1}{12}
\ln \frac{\Lambda^2}{m^2_r} - \frac{5}{72} \right) \\
T & =& \frac{3\gamma^2}{16 \pi \cos^2 \theta_W} \left( -\frac{a_M}{3}+
\ln \frac{\Lambda^2}{m^2_r} +\frac{5}{6} \right).
\end{eqnarray} 
Inspecting the above expressions for the $\Pi$'s one finds that 
there is no divergent contribution to $U$, and this is consistent 
with the fact that ${\cal O}_X$ and ${\cal O}_M$ provide no 
counterterms for $U$. 

It is interesting that the radion contribution to $S$ is 
negative and to $T$ is positive. This is easy to understand 
by comparing this result to the contribution of the Higgs in the SM.
In fact, 
for $\gamma=1$, $m_r=m_h$ and $Z_i(\Lambda)=0$, 
the radion result 
is identical to the (logarithmic) contribution of the SM physical Higgs. 
But there 
 the total correction to $S$ and $T$ 
is finite, so that for the Higgs 
the $\Lambda$ dependence in the above expression 
is canceled by loops of $W$'s and $Z$, leaving a $\ln m_h/M_Z$ 
dependence. Then one obtains the 
usual positive (negative) correction to $S$ $(T)$.
For the radion though the large logarithms are present, 
and since $\Lambda > m_r$, the radion contribution to  
$S$ is negative, and to $T$ it is 
positive. Recalling that 
$\gamma^2 = v^2/ 6 \Lambda^2$,  
the size of the above correction is only significant for small values of 
$\Lambda$.

We conclude with a comment on 
the decoupling behavior of the radion. Inspecting the above 
expressions we see that for large $m_r$ the radion contribution 
scales as 
\begin{equation} 
 \frac{1}{\Lambda^2} \ln \frac{\Lambda_C}{m_r}  .
\end{equation}
For the purposes of this paragraph we distinguish the cutoff 
of the effective theory, $\Lambda_C$, from the mass scale 
$\Lambda = M_{Pl} e^{-k r_0}$ appearing in the radion coupling. 
In our analysis we have approximated $\Lambda_C \sim \Lambda$. 
If the radion mass is much larger then $\Lambda$, then the cutoff 
of the 
effective theory is the much higher 
scale $E \sim m_r$ or larger. Then it is more natural to express all 
higher dimension operators as suppressed by $m^{-1}_r$ or 
$\Lambda_C ^{-1}$. But then 
the coupling of the radion to the gauge bosons contains the very large 
coefficient $m_r / \Lambda$. It is then inappropriate to use the 
one--loop approximation. So to remain within the validity of the 
approximations used here one must also increase $\Lambda$ for large $m_r$. 
Then the radion contribution decouples. 

For large radion mass it is conceivable that the 
coupling of the radion to Tr$T$ decreases. 
This cannot be seen in our computations here because we have made 
the approximation of using only the zero mode wavefunction to determine 
the coupling, but have included the backreaction perturbatively 
to compute the radion mass. For very large radion mass these approximations 
are invalid, and then one must exactly solve the equations. It is then 
conceivable that for large radion mass the radion wavefunction 
on the TeV brane decreases in such a manner that the coupling to 
Tr$T$ remains 
natural, i.e. $O(m^{-1}_r)$. 

Following standard practice
we define  
a reference model in which one computes the oblique 
parameters within the Standard Model, which means 
for some specific value for the 
Higgs mass. Since the curvature scalar operator mixes the 
radion and Higgs, the two physical scalars are some 
mixture of the gauge Higgs and radion. This mixture is 
not a unitary rotation due to the kinetic mixing between the 
states. 
The couplings of the `Higgs' in this case is 
somewhat different than in the Standard Model, and for the purposes 
of computing the oblique parameters it is easiest to think of this 
as a new model, rather than as a perturbation to the Standard Model.
So in computing the oblique parameters the Standard Model 
Higgs contribution (for the reference Higgs mass) 
should be subtracted out, and the contribution 
of the physical states in this model added back in. That is,  
\begin{equation} 
X-X^{ref} _{SM} = X_{new}(m_h,m_r,\Lambda,\xi) - X^{ref}_{H}(m_h=m^{ref}_h) .
\end{equation}
As mentioned, $X^{ref}_{H}(m_h=m^{ref}_h)$ is the contribution from 
only the 
Higgs, with 
mass $m^{ref}_h$ and  with
Standard Model couplings, and it is independent of $\Lambda$, $m_r$ and 
the curvature-mixing parameter $\xi$. The quantity $X^{ref}_{SM}$ is 
the full SM contribution, with the Higgs set at the reference mass 
$m^{ref} _h$. 
The new physics contribution contains two pieces, 
\begin{equation} 
X_{new} = X_R + X_H \hbox{ ,}
\end{equation} 
which describe the contribution of the physical radion $(X_R)$, 
and the physical Higgs $(X_H)$. In the limit of no 
curvature--scalar mixing, this last contribution is just that of 
a Standard Model Higgs with mass $m_h$. For a general curvature 
scalar mixing one just needs to include the effect of the mixing 
coefficients $a$, $b$, $c$ and $d$ given in the previous section. Then 
\begin{eqnarray} 
X_{new} &=& 
\left(\cos \theta - \frac{\gamma}{Z} (6 \xi -1) \sin \theta \right)^2
 X(m^{phys}_h, \gamma=1) \nonumber \\
& &+ 
\left(\sin \theta + \frac{\gamma}{Z} (6 \xi -1) \cos \theta \right)^2
X(m^{phys}_r, \gamma=1) 
\nonumber \\ 
& &- \frac{\gamma^2}{Z^2} (6 \xi-1) X^A . 
\label{generalX}
\end{eqnarray} 
Here $X^A$ is the anomalous contribution, obtained from the vacuum 
polarization (\ref{avac}), but dropping the $m^2_r$ term since this 
does not contribute to any of the oblique parameters. The other
$X$'s are obtained from using the vacuum polarizations (\ref{vp1}),
(\ref{vp2}), 
and inserting the physical mass of the appropriate state. 

We note that, for example, the full anomalous contribution to $S$ from 
the above formula is 
\begin{equation} 
S^A _{new}= \frac{5\gamma^2}{6 Z^2 \pi} (6 \xi-1) 
\end{equation} 
and is negligible for reasonable values of $\gamma$ and $\xi$. 
The anomalous contribution to $T$ is even smaller.

\subsection{Numerical Results}
\label{numerical-sec}

\begin{figure}[t]
\centerline{
\hfill
\epsfxsize=0.60\textwidth
\epsffile{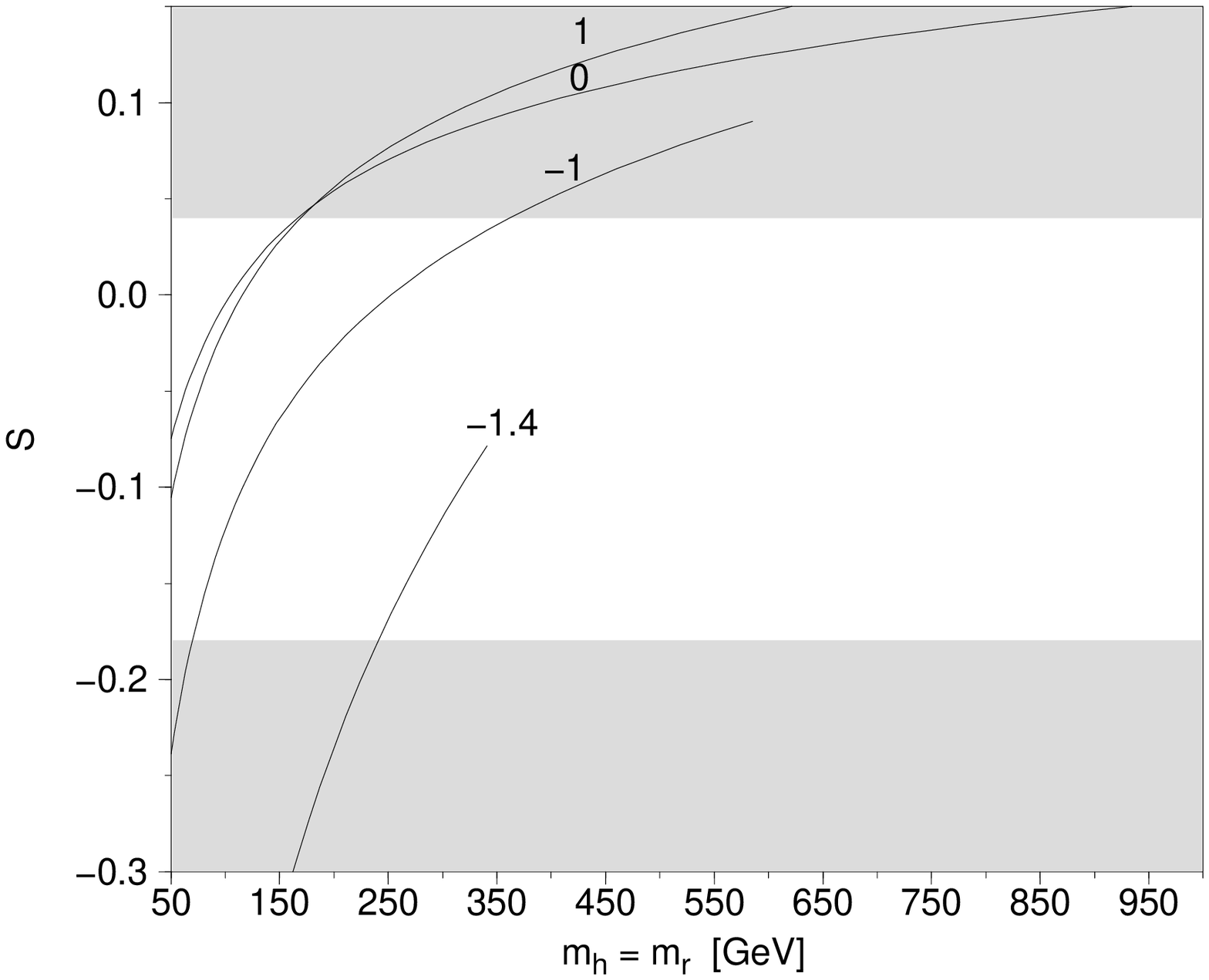}
\hfill
\epsfxsize=0.60\textwidth
\epsffile{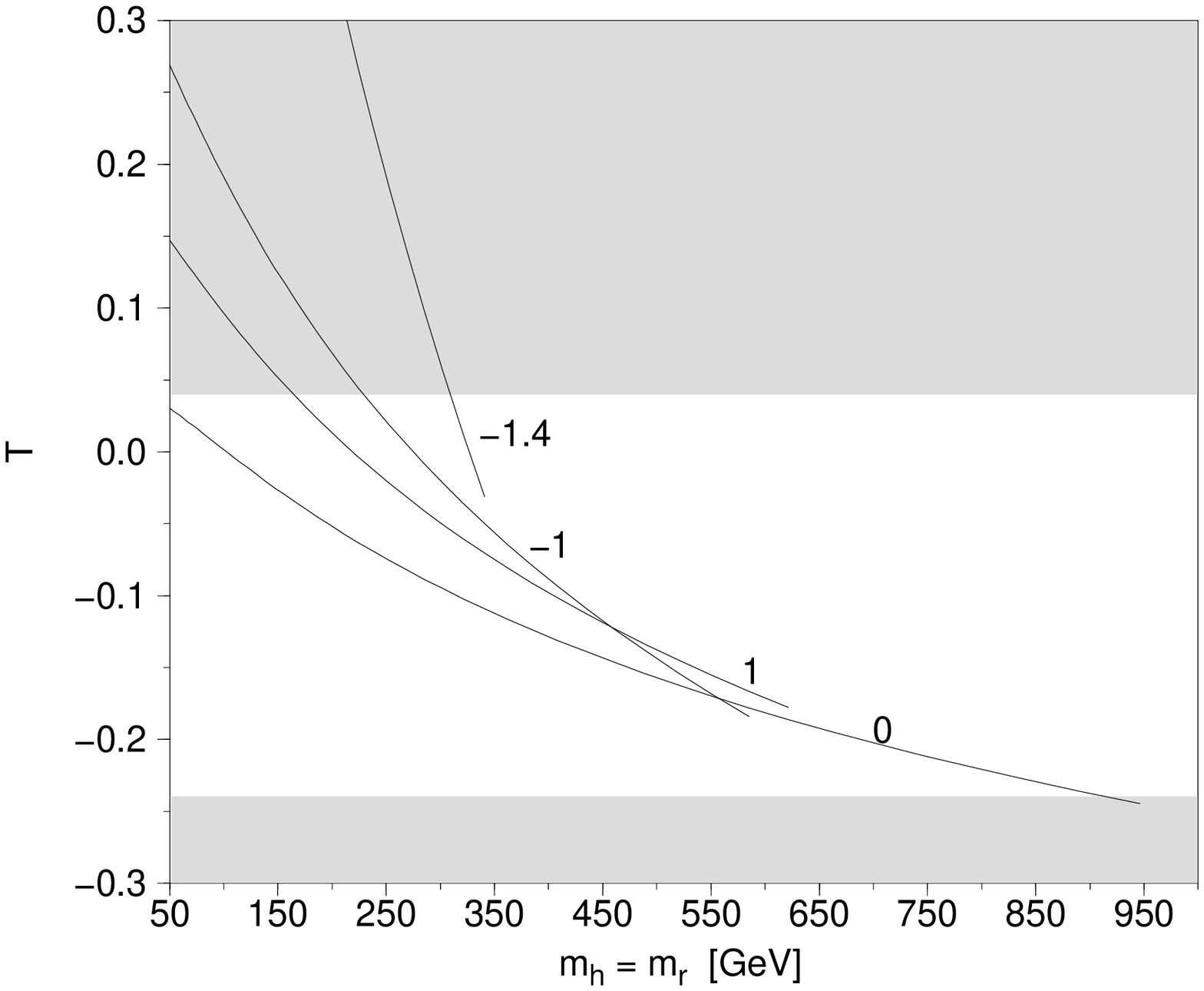}
\hfill }
\caption{The contributions to $S$, $T$ as a function of the
``gauge'' masses $m_h = m_r$.  Each line
is a contour for a fixed curvature scalar mixing $\xi$.
The cutoff scale was chosen to be $1$ TeV ($\gamma = 0.1$).
The shaded regions are excluded by the PDG measurements
to one sigma.}
\label{S-ex-fig}
\end{figure}

\begin{figure}[t]
\centerline{
\hfill
\epsfxsize=0.60\textwidth
\epsffile{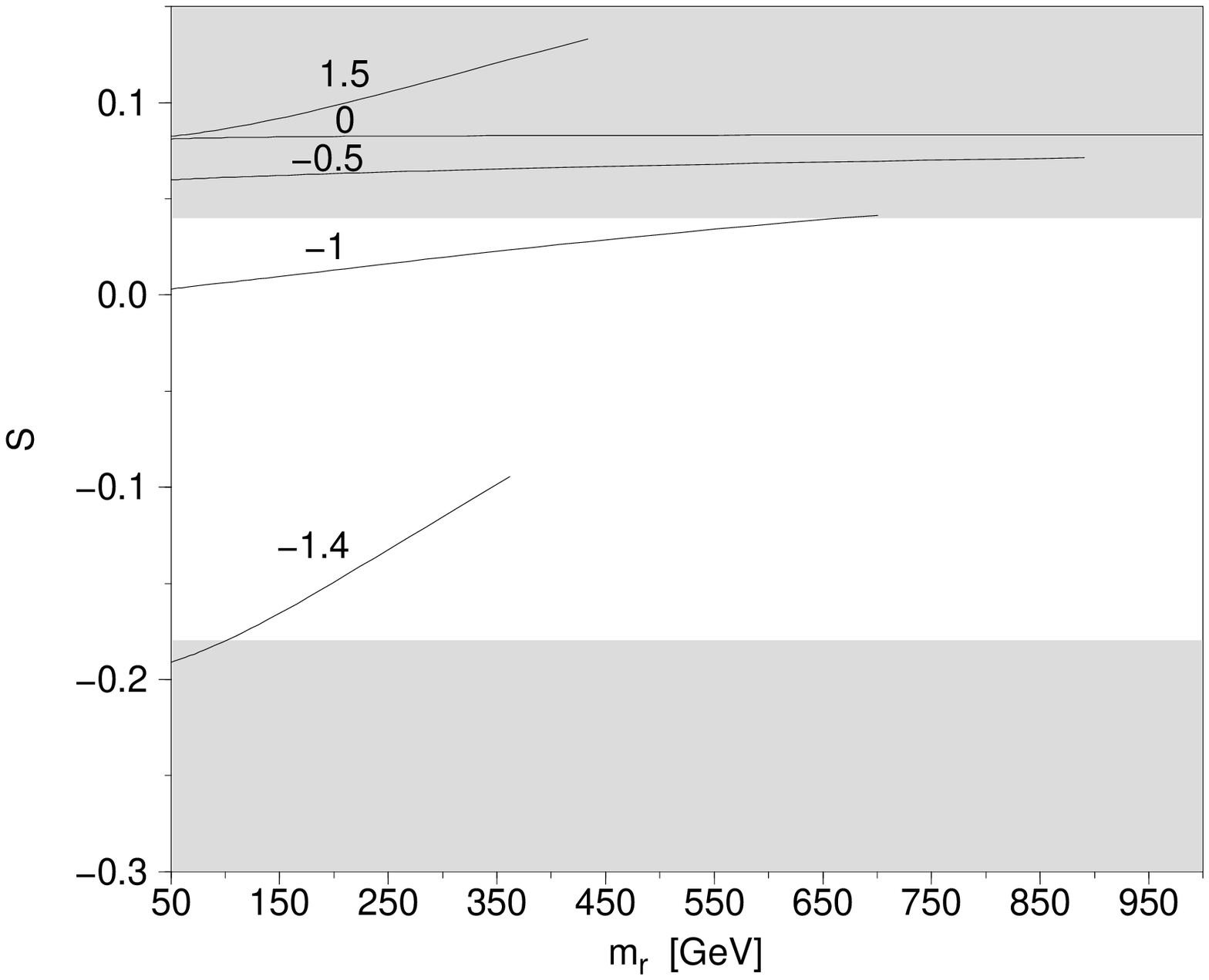}
\hfill
\epsfxsize=0.60\textwidth
\epsffile{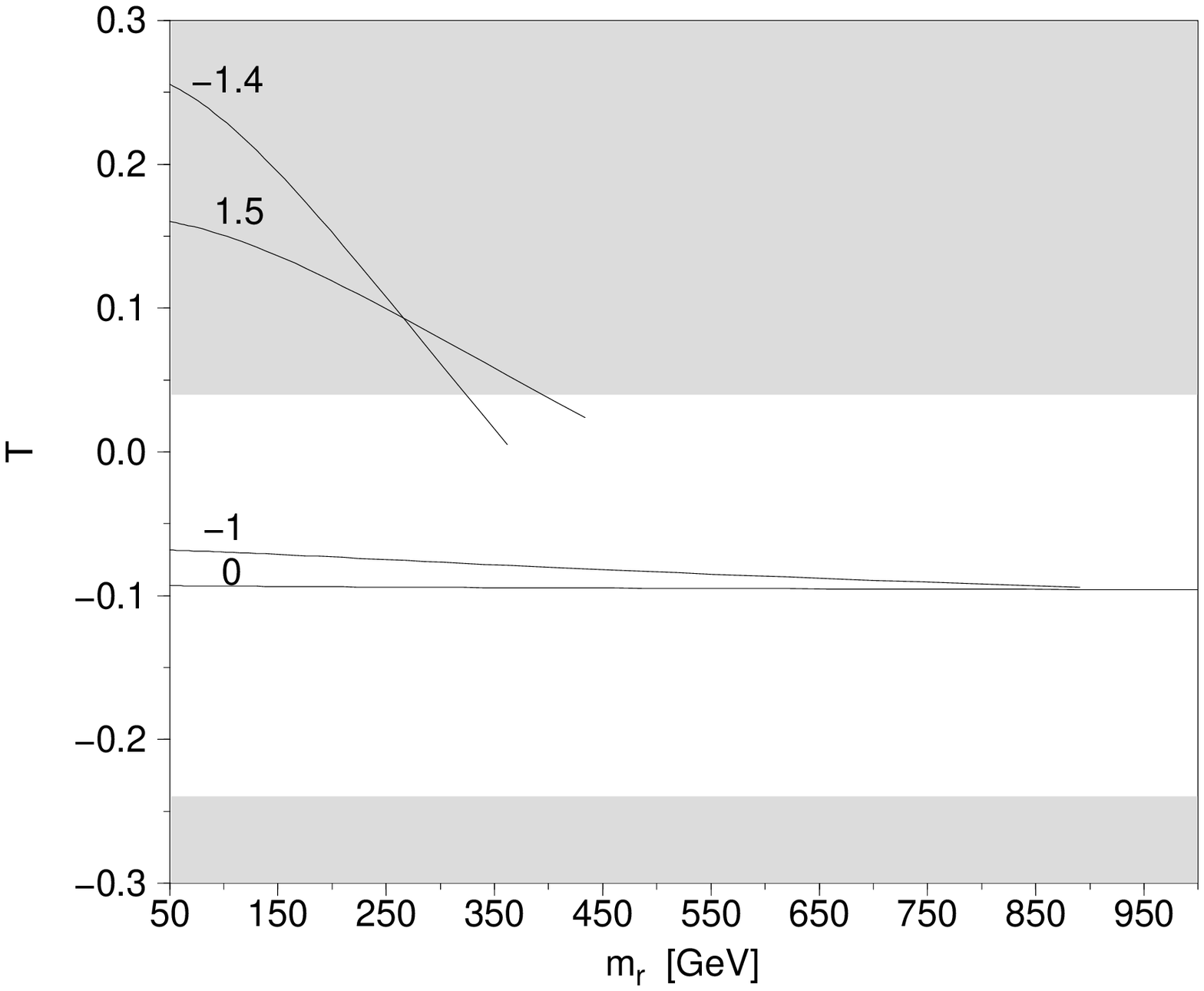}
\hfill }
\caption{Same as Fig.~\ref{S-ex-fig} except that $m_h$ is fixed
to 300 GeV.  Notice that the contributions to $S$ and $T$ are
nearly independent of the radion mass if the curvature mixing
is small since the radion contribution is suppressed by $\sim \gamma$.}
\label{S-mhfix-fig}
\end{figure}

The ``new'' contribution, $X$, is constrained to lie within the measured 
values (extracted assuming $m_h^{\rm SM~ref} = 100$ GeV) \cite{PDG}
\begin{eqnarray}
S_{\rm meas} &=& -0.07 \pm 0.11 \\
T_{\rm meas} &=& -0.10 \pm 0.14 \\
U_{\rm meas} &=& 0.11 \pm 0.15 \; .
\end{eqnarray}
(The errors are for one sigma.)
$X$ can be easily calculated once $m_h$, $m_r$, $\Lambda$, and $\xi$
are specified.  Notice that since the current best-fit values for 
the electroweak parameters are nonzero, the new contributions 
can be more weakly or more strongly constrained
depending on whether they add destructively or constructively 
with the Higgs, respectively.
As a first example, we show the contribution to 
$S$ and $T$ in Fig.~\ref{S-ex-fig} 
as a function of 
$m_h = m_r$ (the ``gauge'' masses), fixing
$\Lambda = 1$ TeV.  Each contour corresponds to a 
different value of $\xi$, and the contours end when a physical mass 
exceeds the cutoff.  
The unshaded region corresponds to the $1\sigma$ allowed
region.  Notice that $T$ is a strong constraint on small (gauge) masses, 
while $S$ is a strong constraint for large masses.  Also, the
the case with $\xi=0$ is nearly identical to the ordinary SM Higgs 
contribution, since the radion contribution that can be 
separated out is strongly suppressed by the coupling $\gamma^2$.
In Fig.~\ref{S-mhfix-fig} we show the contribution to $S$ and $T$
as a function of $m_r$ with $m_h = 300$ GeV.  Notice that the
contributions are nearly independent of $m_r$ for small 
curvature scalar mixing.

\begin{figure}[t]
\centerline{
\hfill
\epsfxsize=0.60\textwidth
\epsffile{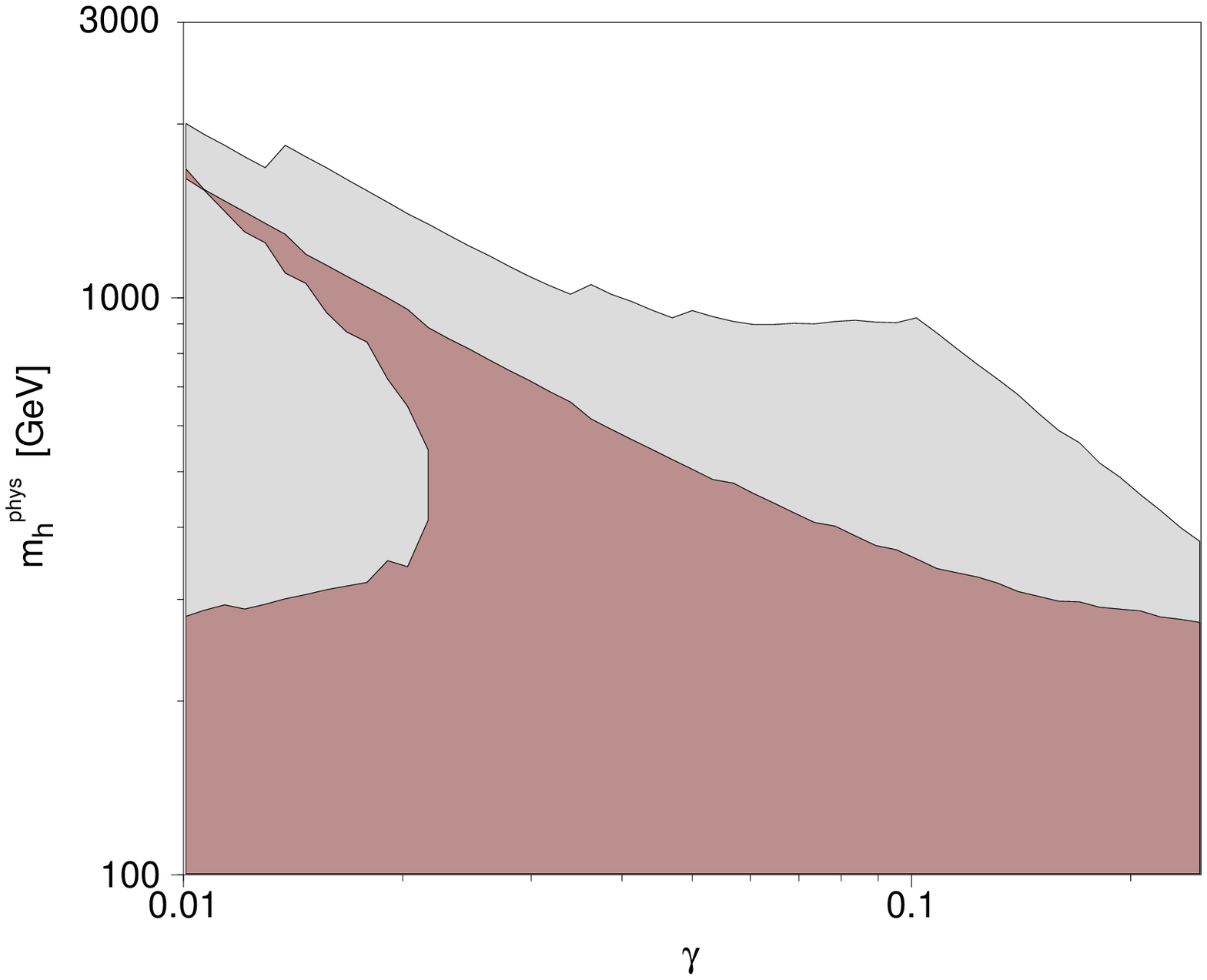}
\hfill
\epsfxsize=0.60\textwidth
\epsffile{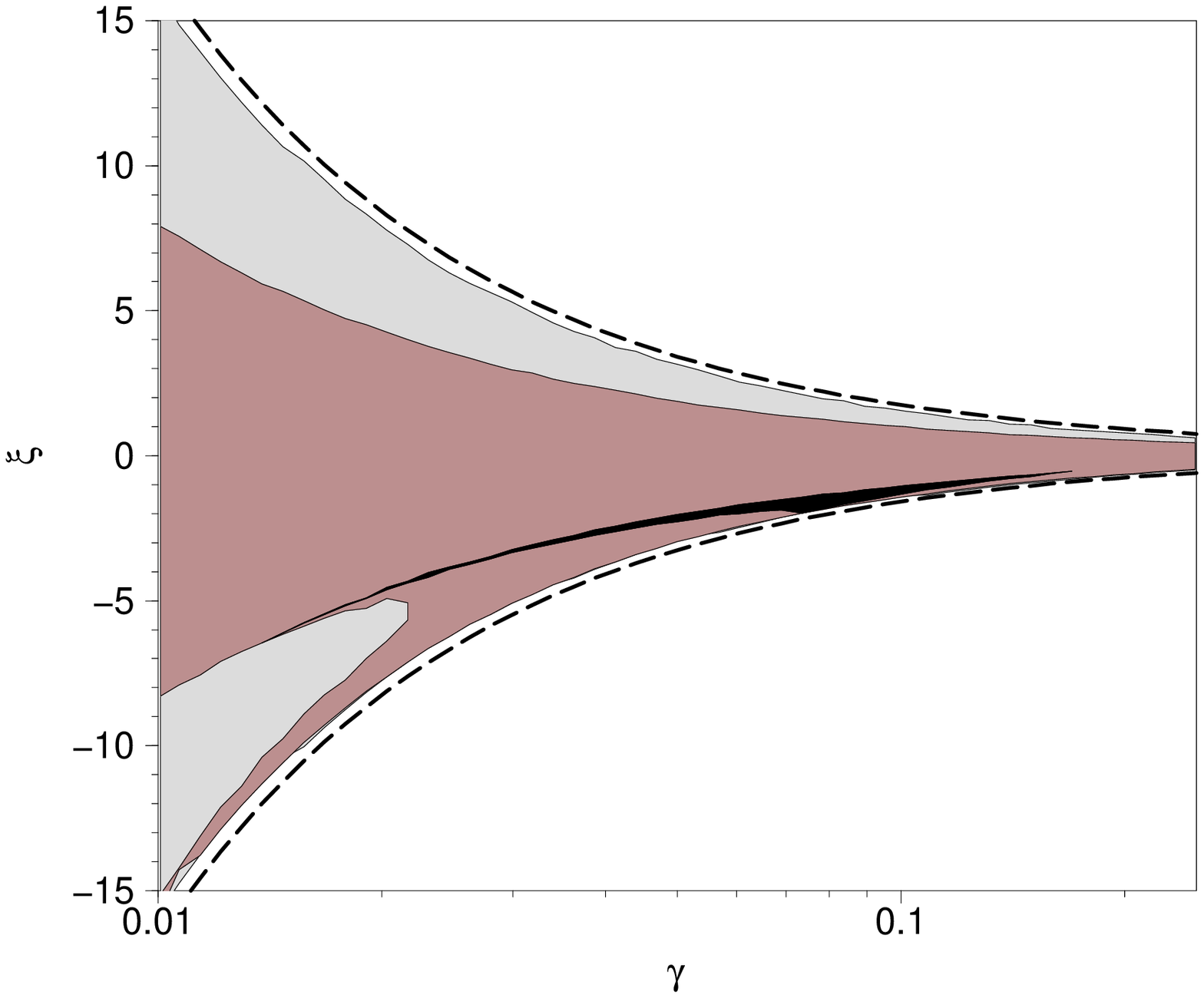}
\hfill }
\caption{The allowed region of $m_h^{\rm phys}$ and $\xi$
as a function of the inverse of the cutoff scale $\gamma = v/\Lambda$
by requiring $S,T,U$ do not exceed the one sigma (dark region)
or two sigma (light region) measurements from the PDG.
The dashed lines correspond to the theoretical bound
requiring the kinetic term is non-negative [see Eq.~(9.14)].
The black sliver corresponds to the region where 
$m_h^{\rm phys} \approx 300$ GeV.}
\label{mphys-space-fig}
\end{figure}

The above results illustrate a general trend that with small 
or absent curvature-scalar mixing, the bound on the Higgs mass 
is not significantly affected by the presence of the radion.
This is not true, however, if we allow $\xi$ to take larger
values.  It is easiest to first illustrate that radion 
physics with large curvature scalar mixing can significantly relax 
the bound on the Higgs mass, by scanning through the parameter
space (choosing $m_h = m_r$ for simplicity) for values
that satisfy one or two sigma limits on the electroweak 
parameters.  We find sets of parameters that are not minor 
perturbations on the SM limit allow the physical masses of the 
Higgs and radion to be several hundred GeV, and perhaps even TeV scale.  
In Fig.~\ref{mphys-space-fig} 
we show the range of physical masses and the range of $\xi$ as a function 
of the cutoff scale.  In general there is not a unique
mapping between the figures, however the ``shark fin'' structure
for the one sigma region in Fig.~\ref{mphys-space-fig}(a) does 
correspond to the ``inverse fin'' in Fig.~
\ref{mphys-space-fig}(b).  Notice that at two sigma the
physical Higgs mass can be much larger than the SM bound
throughout the parameter space shown, and even at one sigma
there exists a narrow range of large, negative curvature-scalar 
mixing where the physical Higgs mass could be of order a TeV.
The latter result arises from a cancellation between the physical
Higgs and radion contributions with the SM reference contribution.
This can be seen in the limit of a large radion and Higgs mass. 
Since the dependence on the masses is only logarithmic, we can 
approximate the masses as being equal. Then, for example, 
\begin{equation} 
S_{\rm new}=\frac{1}{\pi} ( \frac{1}{12} \ln \frac{m^2}{M_Z^2} -\frac{5}{72})-
(6 \xi -1)^2 \frac{\gamma^2}{Z^2 \pi} 
(\frac{1}{12} \ln \frac{\Lambda^2}{m^2} +\frac{5}{72}).
\end{equation}
The first contribution is just the usual correction from the Higgs. 
But the second correction can be potentially large and negative due 
to the dependence on the $\xi$ parameter.
It should be emphasized that the large correction is due to the 
(non--unitary) kinetic mixing between the radion and Higgs, 
or equivalently, due to the non--standard couplings of the 
radion and Higgs in the mass basis.    
Hence, while this region is provocative, 
it nonetheless requires fine-tuning.

\begin{figure}[t]
\centerline{
\hfill
\epsfxsize=0.60\textwidth
\epsffile{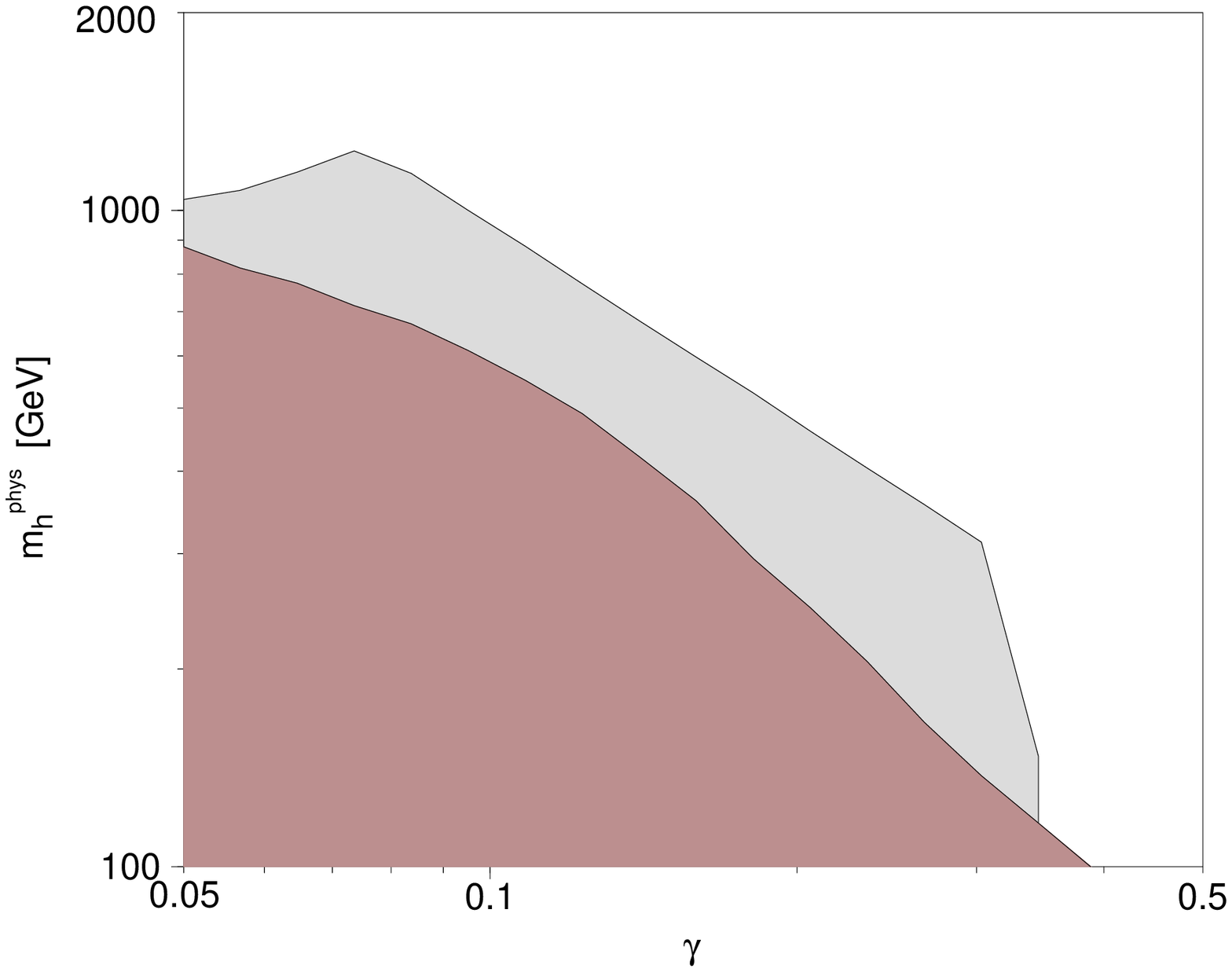}
\hfill
\epsfxsize=0.60\textwidth
\epsffile{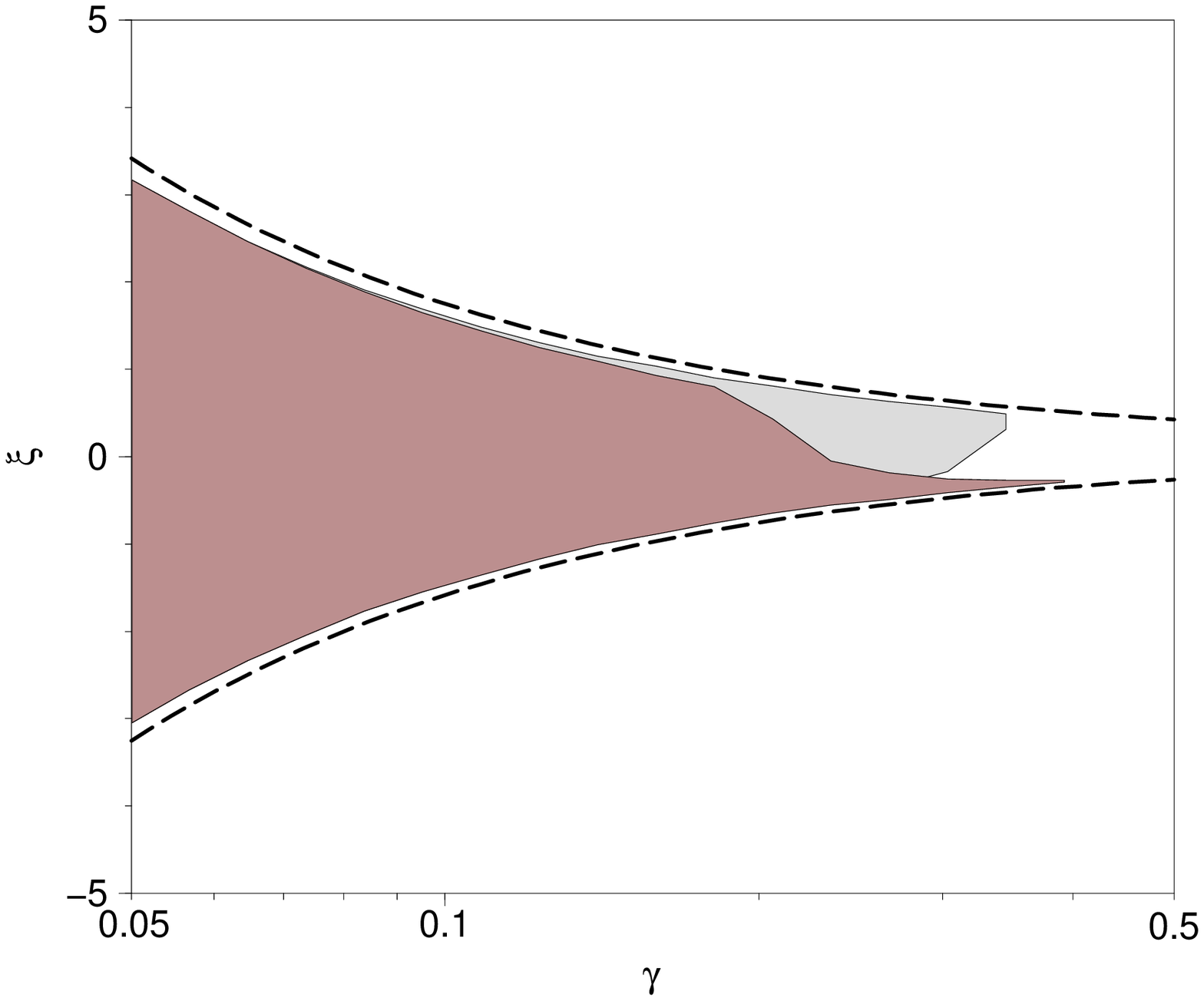}
\hfill }
\caption{The shift in the two sigma contours shown in 
Fig.~\ref{mphys-space-fig} resulting from taking the
coefficients of the nonrenormalizable operators to be
$a_X = 10$ (dark region) and $a_X = -10$ (light region).}
\label{nr-fig}
\end{figure}

These results have assumed that the contribution from the
nonrenormalizable counterterms is small, meaning 
$a_X$ and $a_M$ are less than order $1$.  For larger coefficients 
the allowed regions of parameter space, albeit only
at moderately large $\gamma$.  In Fig.~\ref{nr-fig}
we show the shift in the contours, for the two sigma region,
resulting from taking $a_X = \pm 10$.  ($a_M$ was also
taken to be $10$, but the effect on the contours was
negligible.)  

\section{Limits on Radion Mass}
\setcounter{equation}{0}
\setcounter{footnote}{0}
\label{limits-sec}

As we found in Section \ref{mass-sec}, the mass of the radion is 
expected to be significantly below the the cutoff scale, 
placing it in a region that can be directly probed by experiments.  
The previous section has shown 
that the radion couples much like a Higgs boson, and in the limit 
$\xi \rightarrow 0$, the tree-level couplings of the radion are 
simply scaled by $\gamma$.  Let us first consider the bounds in 
this case.

In the SM, the current bound on the Higgs mass comes primarily
from the LEP processes $e^+e^- \rightarrow Z^* \rightarrow Zh$, with the
value $m_h^{\rm SM} \lsim 108$ GeV \cite{combined-LEP-2000}.
For the radion, an exactly analogous production process occurs
$e^+e^- \rightarrow Z^* \rightarrow Zr$, except that the $ZZr$
coupling has a factor of $\gamma$.  To a good approximation, we can 
therefore estimate the production cross section of radions at LEP
by simply scaling the Higgs cross section by $\gamma^2$.

The decay of the radion is somewhat more complicated, however.
As we discussed in Section \ref{feynman-sec}, the radion couples
directly to gauge bosons through the conformal anomaly.  Although
this coupling is one-loop suppressed, it competes with Yukawa
suppressed interactions and, for the case $rgg$, can be comparable
or even dominate \cite{GRW, Korean}.  In the radion mass range well 
below the $t\overline{t}$ threshold, the ratio of the two largest
widths can be expressed as
\begin{equation}
\frac{\Gamma(r \rightarrow gg)}{\Gamma(r \rightarrow b\overline{b})} = 
\frac{\alpha_s^2 c_3^2}{12 \pi^2 \beta^3} \left( \frac{m_r}{m_b}
\right)^2 \; ,
\end{equation}
where $\beta^2 = 1 - 4 m_b^2/m_r^2$ and $c_3 \approx 23/3$ is roughly
the one-loop QCD $\beta$-function coefficient (approximately including the 
smaller contribution from the one-loop triangle diagram with top quarks.)
Notice that the coupling $\gamma^2$ cancels in this ratio.
This can be written in the suggestive form
\begin{equation}
\frac{\Gamma(r \rightarrow gg)}{\Gamma(r \rightarrow b\overline{b})}
\approx
1/\beta^3 \left( \frac{m_r}{12 m_b} \right)^2 \; .
\end{equation}
Hence, $r \rightarrow gg$ dominates for the region 
$12 m_b \lsim m_r \lsim 2 M_W$.  The search strategy for the radion 
is therefore significantly different from the Higgs in this mass window,
namely searching for a pair of gluon jets instead of a pair of $b$-jets.
Similarly, the radion has a different production cross section at 
hadron colliders via gluon fusion, proportional to the conformal
anomaly-enhanced width into gluons but suppressed by the usual 
$\gamma^2$ \cite{GRW}.

\begin{figure}[t]
\centering
\epsfxsize=4.0in
\hspace*{0in}
\epsffile{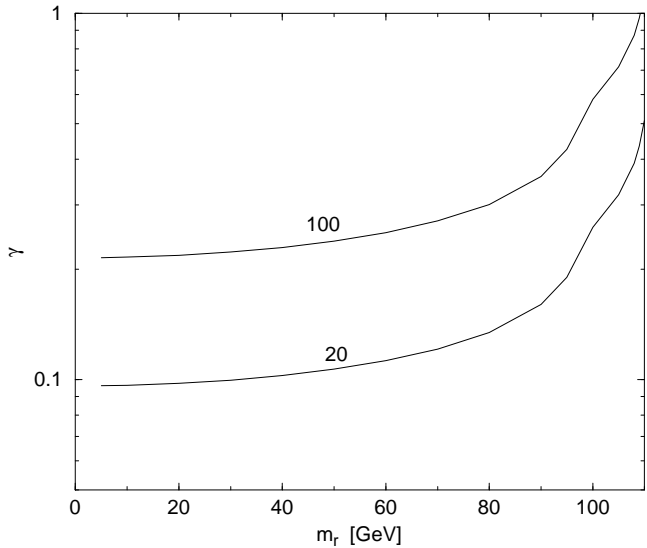}
\caption{The bound on the radion mass as a function of $\gamma$
is shown, assuming the signal could be extracted from background
once the radion production cross section times integrated luminosity 
exceeds $N=20,100$ events at LEP (summed over the four experiments).}
\label{gamma-bound-fig}
\end{figure}

Determining an accurate bound on the radion mass in the region
that can be probed by LEP requires a detailed analysis of detecting 
a two gluon plus Z signal.  We will not attempt this here.  
Instead, the expected bound on the radion mass can be roughly
estimated as a function of the coupling if we assume that some 
number of production events $N$ at LEP could not have escaped detection 
(or be lumped into SM backgrounds).  Near the kinematical
limit, the best bound will always come from the highest energy
data.  For lower mass radions, a lower center-of-mass energy
results in a slightly higher cross section.\footnote{For instance, 
$\sigma_{\sqrt{s} = 189 \; {\rm GeV}}(e^+e^- \rightarrow rZ)/
\sigma_{\sqrt{s} = 202 \; {\rm GeV}}(e^+e^- \rightarrow rZ) \sim 1.3$
for small $m_r$.}  Since the bound for a given radion mass is 
limited only by luminosity, we can combine the multitude
of LEP runs at various energies by weighting by the integrated
luminosity accumulated.  The bound is then simply
\begin{equation}
\gamma^2 < \frac{N}{\sum \sigma_{\sqrt{s}}(e^+e^- \rightarrow Zh; m_h=m_r)
\times \int {\cal L}_{\sqrt{s}}}
\end{equation}
where the sum is over the various recent LEP runs with center-of-mass
energy $\sqrt{s}$.
$N$ encodes all of the detailed analyses of backgrounds, 
signal efficiencies, etc., and is in general not independent of energy 
or radion mass.  In Fig.~\ref{gamma-bound-fig} we simply show the 
bound obtained if $N = 20$ or $100$ (the integrated luminosity
for each energy was also summed) corresponding to producing $20$ or $100$ 
events summed over all four LEP experiments.  These numbers were
chosen since searches for Higgs bosons typically need a few to tens 
of events (per detector) for a statistically significant 
signal-to-background ratio.  Notice that no bound on the radion 
mass is expected from the recent LEP runs once $\gamma$ is less than 
about $0.1$. 

One could also search for light radions, $m_r \lsim 60$ GeV, 
at LEP I via the decay $Z \rightarrow f\overline{f} + r$, 
through the same coupling discussed above.  However, the
expected bound obtained by using this procedure is no better than 
that found above for $m_r$ larger than about $10$ GeV.

We have not attempted to estimate a bound on $\gamma$ for a radion 
mass less than about $10$ GeV.  We really do not expect the 
radion to be several orders of magnitude below the cutoff scale,
and so at the outset it seems this mass region is unnatural.
But, the presence of several low energy production processes 
(and rare decays) could be important, so a considerably more careful 
analysis than what we have attempted here is needed.

When curvature scalar mixing is included, the coupling $ZZr$
is modified as shown in Eq.~(\ref{eff-mixed-lag-eq}).
The above analysis can be translated into this more general case,
but now the coupling is not simply $\gamma$ but a function of
the curvature-scalar mixing as well.  In addition, the SM Higgs 
couplings are also modified, and so its production and decay 
are also affected.  In particular, the production cross section 
could be either enhanced or suppressed.  (This is similar to what 
happens in two Higgs doublet models, such as the MSSM.)
An interesting signal for RS with curvature-scalar mixing
could be observing a nonstandard cross section or decay rate 
for a SM-like Higgs.

\section{Conclusions}
\setcounter{equation}{0}
\setcounter{footnote}{0}
\label{conclusions-sec}

In this paper we have analyzed the coupled radion-scalar system 
in detail, including the backreaction of the bulk stabilizing scalar
on the metric.  We derived the coupled differential equations 
governing the dynamics of the system, and found the mass eigenvalues
for some limiting cases.  We find that due to the coupling between 
the radion and the bulk scalar, there will be a single KK tower describing
the system, with the metric perturbations non-vanishing for every 
KK mode. This implies that the Standard
Model fields localized on the TeV brane will couple to every KK mode from 
the bulk scalar, and this could provide a means to directly 
probe the stabilizing physics. We also found that in an expanding 
universe the shift in the 
radion at late times completely agrees with the effective theory 
result of \cite{CGRT}.
  
We also calculated the contributions of the radion to the oblique 
parameters using an effective theory approach.  Since the radion is 
the only new state well below the TeV scale, we argued that a low--energy 
effective theory including only the radion and SM fields is sufficient, 
as long as appropriate nonrenormalizable counterterms at the cutoff 
scale are added.  In the absence of a curvature--scalar Higgs mixing 
operator, the size of the contribution to the oblique parameters due 
to the radion is small.  In the presence of such a mixing operator, 
the corrections can be much larger due to the modified radion and Higgs 
couplings.  In particular, including only the mixed radion and Higgs 
fields as ``new physics'', we calculated the range of curvature-scalar 
mixing for a given cutoff scale that allows the physical Higgs mass 
to be up to of order the cutoff scale, while $S_{\rm new}$ and $T_{\rm new}$ 
were within the experimental limits.  However, the parameters must be 
increasingly fine-tuned to achieve a Higgs mass that exceeds a few 
hundred GeV.

\section*{Acknowledgements}
\setcounter{equation}{0}
\setcounter{footnote}{0}

After Sections \ref{coupled-sec}-\ref{mass-sec} were completed, 
we were informed that many of the
results of these sections are also contained in Ref.~\cite{TanakaMontes}.
We thank Riccardo Rattazzi for pointing us to this reference.
We also thank Tanmoy Bhattacharya, Kiwoon Choi, Jim Cline, 
G\'abor Cynolter, Josh Erlich, 
Christophe Grojean, Howie Haber, Tao Han, Gy\"orgy P\'ocsik, Lisa Randall,
John Terning and James Wells
for useful discussions, Jim Cline for reading 
an earlier version of the manuscript, as well as Tao Han and John Terning
for comments on the manuscript. 
C.C. is an Oppenheimer fellow at the Los Alamos National Laboratory.
The research of C.C. is supported by
the Department of Energy under contract W-7405-ENG-36.
The work of M.G. was supported in part by the Department 
of Energy.  The research of G.K. is supported in part by
the U.S. Department 
of Energy under grant number DOE-ER-40682-143.

\end{document}